\documentclass{elsart}
\usepackage[pctex32]{graphics}
\usepackage{graphicx,amsmath,verbatim,amssymb,psfrag}

\begin{document}
\begin{frontmatter}
\title{Statistical modelling of higher-order correlations in pools of neural activity}
\author[DtoFis,Iflysib]{Fernando Montani},
\ead{fmontani@gmail.com}
\corauth[cor]{Corresponding author}
\author[ICBioeng,ICMat]{Elena Phoka},
\author[IFLP]{Mariela Portesi},
\author[ICBioeng]{Simon R. Schultz}
\address[DtoFis]{Departamento de F\'{\i}sica, Facultad de Ciencias Exactas, \\
           UNLP Calle 49 y 115. C.C. 67 (1900), La Plata, Argentina.}

\address[Iflysib]{IFLYSIB, Universidad Nacional de La Plata, La Plata, Argentina}

\address[IFLP]{IFLP, Universidad Nacional de La Plata, CC67, 1900 La Plata, Argentina}

\address[ICMat]{Department of Mathematics, Imperial College London, South Kensington, London SW7 2AZ,
United Kingdom.}

\address[ICBioeng]{Department of Bioengineering, Imperial College London, South Kensington, London SW7 2AZ,
United Kingdom}
\begin{abstract}
Simultaneous recordings from multiple neural units allow us to investigate the activity of very large
neural ensembles. To understand how large ensembles of neurons process sensory information,
it is necessary to develop suitable statistical models to describe the response variability of the recorded spike trains.
Using the information geometry framework, it is possible to estimate higher-order correlations by assigning one interaction
parameter to each degree of correlation, leading to a $(2^N-1)$-dimensional model for a population with $N$ neurons.
However, this model suffers greatly from a combinatorial explosion, and the number of parameters to be estimated from the
available sample size constitutes the main intractability reason of this approach. To quantify the extent of higher than pairwise spike correlations in pools of multiunit activity, we use an information-geometric approach within the framework of the extended
central limit theorem considering all possible contributions from high-order spike correlations.
The identification of a deformation parameter allows us to provide a statistical characterisation
of the amount of high-order correlations in the case of a very large neural ensemble, significantly
reducing the number of parameters, avoiding the sampling problem, and inferring the underlying dynamical
properties of the network within pools of multiunit neural activity.
\newenvironment{keywords}{
       \list{}{\advance\topsep by0.35cm\relax\small
       \leftmargin=1cm
       \labelwidth=0.35cm
       \listparindent=0.35cm
       \itemindent\listparindent
       \rightmargin\leftmargin}\item[\hskip\labelsep
                                     \bfseries Keywords:]}
     {\endlist}
\end{abstract}
\maketitle
\end{frontmatter}

\section{Introduction}
\label{sec:Intro}
To understand how sensory information is processed in the brain, we
need to investigate how information is represented collectively by the
activity of a population of neurons. There is a large body of evidence
suggesting that pairwise correlations are important for information
representation or processing in retina
\cite{Schneidman_Nature_06,Shlens06}, thalamus \cite{Alonso08} and
cerebral \cite{KohnSmith05,SamondsBonds05,Montani07} and
cerebellar cortices \cite{Schultz09,Wise10}.
However, there is also evidence that in at least some circumstances, pairwise correlations
do not by themselves account for multineuronal firing patterns \cite{Montani09,victor10}; in
such circumstances, triplet and higher-order interactions are important. The role of such higher-order
interactions in information processing
is still to be determined, although they are often interpreted as a signature of the formation of Hebbian cell assemblies
\cite{Hebb49}.

Higher-order correlations may be important for neural coding even if
they arise from random fluctuations. Amari and colleagues
\cite{AmariWu} have suggested that a
widespread distribution of neuronal activity can generate high-order stochastic interactions. In
this case, pairwise or even triplet-wise correlations do not uniquely determine
synchronised spiking in a population of neurons, and high-order
interactions across neurons cannot be disregarded. Thus, to gain a
better understanding of how neural information is processed, we
need to study, whether higher-order interactions arise from cell
assemblies or from stochastic fluctuations. Information-geometric
measures can be used to analyse neural firing patterns including
correlations of all orders across the population of neurons
\cite{Bohte00,Amari2001,Naka02,Tanaka,Ikeda04,Wu04,Montani09,victor10}.

A straightforward way to investigate the neural activity of a large population of neurons
is to use binary maximum entropy models incorporating pairwise correlations on short time scales \cite{Schneidman_Nature_06,Shlens06}. To estimate this model, one has to consider a sufficient amount of data
to measure the mean activity of individual neurons and correlations in pairs of neurons. This allows us to estimate the functional connectivity in a population of neurons at pairwise level. However, if higher-order correlations are present in the data, and as the number of possible binary
patterns grows exponentially with the number of neurons, we would need to use an appropriate
mathematical approach to go beyond a pairwise modelling.
This is in general a difficult problem, as sampling even third-order interactions can be
difficult in a real neurophysiological setting \cite{Montani09,victor10}.

However, under particular constraints, this sampling difficulty can be
substantially ameliorated. An example of such a constraint is {\em pooling},
in which the identity of which neuron fires a spike in
each pool is disregarded. Such a pooling process reflects the
behaviour of a simple integrate-and-fire neuron model in reading out
the activity of an ensemble of neurons. It also reflects a common
measurement made in systems neuroscience: the recording of multiunit
neural activity, without spike sorting. Whether or not such
constraints permit a complete description of neural information
processing is still a matter of debate \cite{Montani07,victor10},
but they may allow substantial insight to be gained into the
mechanisms of information processing in neural circuits. A method for the
statistical quantification of correlations higher than two,
in the representation of information by a neuronal pool, would be
extremely useful, as it would allow the degree of higher - order correlations
to be estimated from recordings of multiunit activity (MUA) - which can be performed in a much broader range of
circumstances than ``spike-sortable'' recordings can.

In this paper, we use a pooling assumption to investigate
the limit of a very large neural ensemble, within the framework of information
geometry \cite{AmariWu,Bohte00,Amari2001,Naka02,Tanaka,Ikeda04,Wu04,Amari:80Bib,AmariMono}.
In particular, we take advantage
of the  recent mathematical developments in $q$-geometry (and $q$-information geometry)
and the extended central limit theorem \cite{Amariqexp,umarov1,umarov2,vignat,vignat1,vignat2} to provide a statistical quantification of higher-order spike correlations.
This approach allows us to identify a deformation parameter
to characterise the extent of spike correlations higher than two in the limit
of a very large neuronal ensemble. Our method
accounts for the different regimes of firing within the probability
distribution and provides a phenomenological description of the
data, inferring the underlying role of noise correlations
within pools of multiunit neural activity.

\section{Methodology}
\subsection{Information geometry and the pooled model}
\label{sec:poolmodel}

\noindent We represent neuronal firing in a population of size $N$ by a binary vector
\begin{math} X = (x_1,\ldots,x_N ) \end{math},
where $x_{i}=0$ if neuron $i$ is silent in some time window $\Delta T$
and $x_{i}=1$ if it is firing (see {Fig~\ref{inf00}}{\emph{a}}).  Then, for a given time window, we
consider the probability distribution of binary vectors, $\{P(X)\}$.
Any such probability distribution $\{P(X)\}$ is made up of $2^N$ probabilities
\begin{eqnarray}
P(X)=Prob\{X_1=i_1, ...,X_N=i_N \}=P_{i_1 ... i_N}
\end{eqnarray}
subject to the normalization \begin{equation}
\sum_{i_1,...,i_N=0,1} P_{i_1 ... i_N}=1.\end{equation}

As proposed by Amari and co-workers (see, for instance, Refs \cite{AmariMono,Naka02}),
the set of all the probability distributions $\{P(X) \}$ forms a ($2^N - 1$)-dimensional manifold ${S_N}$.
This approach uses the orthogonality of the
natural and expectation parameters in the exponential family of distributions. It is also useful for analysing neural firing in a systematic manner based on information geometry.
Any such probability distribution can be unequivocally determined using a \textit{coordinate system}.
One possible coordinate system is given by the set of $2^N-1$ marginal probability values:
\begin{eqnarray}
\tilde\eta_i = E[x_i] = P\{ x_i=1  \},   &&   i=1,..., N  \\
\tilde\eta_{ij} = E[x_i x_j] = P\{ x_i=x_j=1  \} , && i<j \\
\vdots \nonumber \\
\tilde\eta_{123...N} = E[x_1 ... x_N] = P\{ x_1=x_2=...=x_N=1 \}.
\end{eqnarray}

These are called the
$\tilde\eta$-coordinates \cite{AmariMono}. Moreover, provided $P(X) \neq
0$, any such distribution can be expanded as in Ref. \cite{Naka02}
\begin{eqnarray}
\lefteqn{
\label{eq:pthetafull}
P(X) = \exp \left\{  \sum_{i=1}^N {x_i } \tilde\theta _i  +
\sum\limits_{i < j} {x_i } x_j \tilde\theta _{ij}  +
\right. } \\
& & \left. + \sum\limits_{i < j < k} {x_i } x_j x_k \tilde\theta _{ijk}  +
\ldots
   + x_1 \ldots x_N \tilde\theta_{1\ldots N}  - \tilde\psi   \right\},
   \nonumber
\end{eqnarray}
where there are in total $2^N-1$ different $\tilde\theta$ correlation coefficients that can be used to
determine univocally the probability distribution. The $\tilde\theta$ forms a coordinate system, named $\tilde\theta-coordinates$,
which correspond to e-flat structure in the $(2^N-1)$-dimensional manifold $S_N$. In Eq.~(\ref{eq:pthetafull}), $\tilde\psi$ is a
normalization term. The $\tilde\eta$-coordinates and $\tilde\theta$ coordinates are dually orthogonal coordinates.
The properties of the dual orthogonal coordinates allow the formulation of the generalised Pythagoras theorem that
gives a decomposition of the Kullback Leibler
divergence to calculate contributions of different orders of interaction between two probability distributions.

These notions are rigorously developed in Refs. \cite{AmariWu,Amari2001,Naka02,Tanaka,Ikeda04,Wu04,Amari:80Bib,AmariMono} within
the framework of information geometry. Assigning one ``interaction parameter'' $\tilde\theta$ to each degree
of correlation, we have a ($2^N-1$)-dimensional
model for a population with $N$ neurons. The basis of this formalism
is shown in {Fig~\ref{inf00}}{\emph{b}}: once our coordinate $\tilde \theta$ is fixed
we have a subset $E(\tilde\theta)$ for each possible value of $\tilde\theta$ and an exponential family of distributions
(with the same value of $\tilde\theta $). Notice that when higher-order correlations are considered, we need to construct an orthogonal
multidimensional coordinate space to build the probability distribution, as is schematised in
{Fig~\ref{inf00}}{\emph{b}}. Let us consider $M(\tilde\eta_1, ...,\tilde\eta_N)$ to be a submanifold of the marginals in
$S_N$. Then, the tangential direction of $M(\tilde\eta_1,...,\tilde\eta_N)$  represents the direction in which
only the pure correlation changes, while the tangential directions of $E(\tilde\theta)$
span the directions in which only $\tilde\eta_1, ...,\tilde\eta_N$ change but $\tilde\theta$ is fixed.
Directions of changes in the correlations and marginals are required to be mutually orthogonal.

It is important to point out that the estimation of all the
parameters associated with higher-order correlations suffers greatly from a combinatorial explosion. Consider, for instance,
a population of 50 neurons for which we would
need therefore more than $10^{15}$ parameters. Thus, the number of parameters to be estimated from the available
sample size constitutes the reason for the intractability of this approach.

However, if we assume that a neuron cannot process spikes from different
neurons separately, then the labels of the neurons that fired each of
the spikes are lost. In this case, the target neuron is only aware of
the {\it number} of synchronous firings among its inputs. Similarly, a
neurophysiological recording technique that disregards the identity
of the neuron that fires each spike (i.e. it measures MUA) can only count
the number of cells in the vicinity firing in a time window, rather than provide
information about their pattern.

To investigate the effects of such
processes, we consider a \textit{pooled model} \cite{Bohte00,AmariWu},
where we assume (for mathematical convenience) a population of $N$
identical neurons.  Rather than the full distribution of $X$ with
probabilities $P(X)$ considered above, we now introduce a set of probabilities $P(k)$ where
$k=0,1,\ldots,N$ represents the number of synchronous spiking neurons
in the population during a time interval $\Delta T$.

This assumption greatly simplifies the analysis based on the
coordinate systems described above. The probability distribution is
now characterised by only $N$ parameters. Following Ref. \cite{Montani09}, we refer to the probability values $P(k)$ as
the $p$-coordinates (the \textit{pooled model} is defined by $N$ parameters). Then, considering
the number of neurons that are firing simultaneously (within $\Delta T$) in the
pooled model, instead of the probability distribution Eq.~\eqref{eq:pthetafull},
one has
\begin{equation}
P(1) = N \exp{\left( \theta_1 - \psi(\theta ) \right)} , \label{opr1} \\
\end{equation}
and
\begin{eqnarray}
P(k) = \binom{N}{k} \,
  \exp{\left( k \ \theta_1  +
  \sum\limits_{i = 2}^k \binom{k}{i} \ \theta_i
  - \psi (\theta ) \right)} , && k=2,\ldots,N
  \label{eqPkpool}
\end{eqnarray}
while
\begin{math}
P(0) = \exp(- \psi(\theta ) )    
\end{math}
corresponds to a completely silent response and is written in such way so as to fulfil normalization
of the probabilities $\{ P(k) \}$.
The marginals $\eta_{k}$, indicating the probability of any $k$
neurons firing within $\Delta T$, are given by
\begin{math}
  \eta_{k} = \sum \limits_{i=k}^{N} \binom{N-k}{i-k} \frac{P(i)}{\binom{N}{i}} ,
\end{math}
with $k=0,1,\ldots,N$ \cite{Bohte00}.
In Eqs.~\eqref{opr1} and \eqref{eqPkpool}, $\theta_1$ corresponds to the first order contribution to the log probabilities, while each $\theta_{i}$ ($i=2,\ldots,N$) represents
the effect on the log probabilities of interactions of order $i$ in the
neuronal pool.  We will deal with the relative probability of
simultaneous spiking.  The ratio between
$P(k)$ and $P(k-1)$ is given by the following expressions: 

\begin{equation} \frac{{P(1)}}{{P(0)}}= N  \exp( \theta_1), \label{eq:frac} \end{equation}  \begin{equation} \frac{{P(2)}}{{P(1)}}= \frac{N-1}{2} \exp( \theta_1 + \theta_2), \label{eq:fracs} \end{equation}
and if $k >2$
\begin{eqnarray}
\frac{{P(k)}}{{P(k - 1)}} = \frac{{N - k + 1}}{k} \, \exp \left(
{\theta _1  + \sum\limits_{i = 2}^{k-1} {\frac{{(k - 1)!}}{{(i-1)!(k- i)!}}}  \ \theta _i
+ \theta_k} \right).
\label{eq:fracbig}
\end{eqnarray}
This approach has been applied in a previous publication \cite{Montani09} within the maximum entropy (ME) principle
to evaluate whether high-order interactions play a role in shaping the
dynamics of neural networks. By fixing the correlation coordinate $\theta_{k+1} = 0$ in Amari's formalism, the ME constraints are set up to order $k$.
This is, the ME principle is archived by simple fixing $\{ \theta_{i} \}_{i > k} = 0$ and the distribution is
determined by $\{ \theta_{i} \}_{i \leq k}$. The constraints of the ME principle allow us to choose the most random distribution
subject to these constraints \cite{Naka02,Montani09}. Thus correlation structures that do not belong to the constrained
features are simply removed. Let us consider for instance ME constraints at pairwise level that are set
by fixing $\{ \theta_{i} \}_{i > 2}= 0$, therefore all correlation structures of order higher than two
are removed.

In a recent paper, Yu et al \cite{plenz} have shown that higher - order correlation structures are quite crucial to get a better understanding of how information might be transmitted in the brain. They have reported the importance of higher - order interactions
to characterise the cortical activity and the dynamics of neural avalanches, exhaustively
analysing how the Ising like models \cite{tkacik,Schneidman_Nature_06,Shlens06,Shlens09} do not provide in general
a fair description of the overall organization of neural interactions when considering large neuronal systems.

\subsection{A simple binomial approach}

In order to get a better understanding of how to take the limit of a very large number of neurons,
we first discuss a quite naive approach using Bernoulli trials and a binomial distribution. Despite its simplicity and limitations,
it helps to provide some intuitive feedback.

 Let us consider the activity of a neuronal
population of $N$ neurons in a specific time window $\Delta T$. It should be noted that $P(k)$
denotes the probability of $k$ cells firing simultaneously and $N-k$ being silent, with $0
\leq k \leq N$. Let us denote by $\lambda$ the mean number of spiking
cells, then $\lambda=E_P[k]=\sum_{k=0}^{N} k P(k) \equiv N \, p$ where we introduced $p=\frac{\lambda}{N}$.

If we assume that members of a population of neurons spike independently
and with a fixed, constant probability, then the spiking activity
of the population can be modelled as Bernoulli trials.
For a population containing $N$ neurons, each with a firing probability
$p$, the probability $P(k)$ of having $k$ neurons firing simultaneously is
given by the binomial distribution:
\begin{math}
  P(k) = b(k;N,p) = \binom{N}{k} p^k (1-p)^{N-k} ,
\end{math}
with $k=0,1,\ldots,N$.
In the limit of large $N$ and small $p$, such that $\lambda=Np$ is finite,
one has the so-called \emph{Poisson regime} \cite{Feller1950}.
In this case, the following relations hold for $k \geq 1$:
\begin{eqnarray}
\frac{{b(k;N,p)}}{{b(k - 1;N,p)}} = \frac{\lambda }{k} +
O(\frac{1}{N}) , \label{fellereq}
\end{eqnarray}
in the particular case $k=1$ we have
\begin{equation}
\frac{{b(1;N,p)}}{{b(0;N,p)}} = \lambda  +
O(\frac{1}{N}) ,
\label{quotb}
\end{equation}
while for $k=0$ the distribution in the \emph{Poisson regime} reads
\begin{equation}
{{b(0;N,p)}} = \exp{(-\lambda) } + O(\frac{1}{N})
\label{lakm}.
\end{equation}
We now introduce these ans\"{a}tze in Eqs.~\eqref{eq:frac}-\eqref{eq:fracbig}.
The ratio between the distribution one would derive in the absence of
knowledge of spike correlation, \textit{$P(1)$}, and the distribution
one would obtain when no spike is being fired, \textit{$P(0)$}, is given by
\begin{math} \frac{P(1)}{P(0)} = N e^{\theta_{1}} .\end{math}
Let us consider the limit \begin{math}N \to \infty \end{math} with
$\lambda=N \, p$ finite ($p$ is a small quantity). Then, comparing
Eq.~(\ref{quotb}) with Eq.~(\ref{eq:frac}), we can write \begin{math}
  \lambda = N  e^{\theta _1 } + O(\frac{1}{N}) ,
\end{math}
which leads to
\begin{eqnarray}
\theta_1 & \approx  & \ln \left( \frac{\lambda}{N} \right).
\label{eq:theta1poisson}
\end{eqnarray}
Notice that in this regime, using Eq.~\eqref{lakm}, the probability distribution $P(k)$ at
$k=0$ behaves as \begin{math}{\rm{\emph{P(0)}}} \equiv e^{-\psi}
  = e^{ - \lambda } + O(\frac{1}{N}) \end{math}, which gives $\psi$ approximately equal to $\lambda$ for the normalizing factor.

Introducing Eq.~(\ref{eq:theta1poisson}) into the probability of $k$ neurons firing, Eq.\eqref{eqPkpool},
as \\ \begin{math}\mathop {\lim }\limits_{N \to \infty } \frac{{N(N -
  1)\ldots(N - (k + 1))}}{{(N)^k }} = 1\end{math}, in the large $N$ limit, we get
\begin{math}
P(k,\lambda) \approx \frac{{\lambda ^k }}{{k!}}e^{-\psi(\theta)}  e^{\
{\sum\limits_{i = 2}^k { \binom{k}{i} }  \theta _i } }.
\end{math}
Summing up, if a neuronal population behaves in the large $N$ limit such that
\begin{equation}
  \psi(\theta) \approx \lambda \label{hip1} \end{equation}
and accomplishes Eq.~(\ref{eq:theta1poisson})
then the probability of $k$ cells firing simultaneously can be expressed as a $\tilde{\delta}$-deformed
Poisson distribution such as:
\begin{eqnarray}
P_{\tilde{\delta}}(k,\lambda) \approx \frac{\lambda^{k} e^{-\lambda} e^{\tilde{\delta}(k)} }{k!},
\label{eq:defPq}
\label{eq-qdef}
\end{eqnarray}
where $\tilde{\delta}(k) \equiv {\sum\limits_{i = 2}^k  \binom{k}{i} \theta
_i }$ is a deformation term that summarises correlations of all orders.

Eq.~(\ref{eq-qdef}) corresponds to a Poisson distribution modulated by
a multiple-correlation : $e^{\tilde{\delta}(k)}$. Notice that in the analytical approach presented above
we have taken the limit of $N \rightarrow \infty$, but if only correlations up to order two dominated the
neuronal spiking, the departure from a Poissonian behaviour in the probability of $k$ ($\geq 2$) firing neurons
would be small. Indeed, applying Amari's prescription \cite{AmariWu}, we obtain
$\tilde{\delta}(2)=  \theta_2 =  O(\frac{1}{N})$ and then $e^{\tilde{\delta}(k)}$ is almost unity. But our derivation
comprises all orders through the factor $e^{\tilde{\delta}(k)}$.

It is important to note that we consider the mean number of spiking cells to be equal to the normalization factor (see Eq.~(\ref{hip1})). This assumption represents only a possible subset of the probability distributions $\{P(X) \}$  presented in Eq.~(\ref{eq:pthetafull}). Thus, in the limit of a very large number of neurons, and if higher - orders are taken into account, we cannot provide an analytical estimate of the entire set of all possible widespread distributions using the  approach of Eq.~(\ref{eq-qdef}).

More importantly, due to the high dimensionality involved in
determining Eq.~(\ref{eq:pthetafull}), or even in our approach of Eq.~(\ref{eq-qdef}), we have to take just a few correlation
terms to avoid having an intractable number of parameters
\cite{Amari2001,Montani09}.

But the more basic limitation of the above approach is due to the fact that taking a limit of a very large number of neurons or ``thermodynamic limit'' essentially means to account for the central limit theorem of statistics (CLT). The CLT articulates conditions under which the mean of a sufficiently large number of independent random variables, each with finite mean and variance, can be considered as normally distributed \cite{sheldon}.
In the next sections we discuss how
to take the limit of a very large number of neurons accounting also for
higher-order correlations in the probability distribution. Importantly, we consider the case of any probability distribution of firing, and do not just simply model the neural activity
through a binomial distribution (or Bernoulli trials).  We make use
of an extension of the central limit theorem that
also accounts for the case of correlated random variables \cite{vignat}.
This is particularly important since it allows us to
extend analytical models accounting for effect the of high - order
correlations on the PDFs, making it possible to compute the emergent
properties of the system in the ``thermodynamic limit''.

\subsection{$q$-Information geometry}

Recent studies on information geometry and complex systems have found a big number of distributions that in the asymptotic
limit obey the power law rather than the Gibbs distribution \cite{AmariMono,TsallisGell-Mann,AmariWu,umarov2,Stumpf}.
These power law distributions seem to rule the asymptotic regime, and the $q$-exponential
distributions used in non-extensive statistical mechanics are very useful for capturing such phenomena
\cite{TsallisGell-Mann,vignat,umarov1,umarov2,vignat1,vignat2,AmariMono,Amariqexp}.
The geometrical structure of such probability distributions is termed $q$-geometry, and its mathematical foundations
have been developed by Tsallis, Gell-Mann, Amari, and collaborators. They have in particular proved that
the $q$-geometrical structure is very important to investigate systems with weakly and strongly correlated random variables
\cite{TsallisGell-Mann,vignat,umarov1,umarov2,vignat1,vignat2,AmariMono,Amariqexp}.

In particular, recently Amari and Ohara \cite{Amariqexp} proved that it is possible to generalise
the $q$-structure to any family of probability of distributions, and that the family of $q$-structure is ubiquitous since the of family of all
probability distributions can always be endowed with the structure of the $q$-exponential family for an arbitrary $q$.

The $q$-exponential is defined as
\begin{equation}
\exp_{q} (x) =
\begin{cases}
[1 + (1-q) (x)]^{\frac{1}{1-q}} & \text{if  $(1 + (1-q) (x) ) > 0$;}\\
0 & \text{otherwise}
\end{cases}
\end{equation}
where the limiting case of $q \rightarrow 1 $ reduces to $\exp_{1} (x) = \exp (x)$. The
$q$-exponential family is defined by generalising Eq.~(\ref{eq:pthetafull}) (for a review see \cite{Amariqexp}) as
\begin{equation}
P_q(x,\Theta) = \exp_q \{ \Theta \cdot x - \psi(\Theta ) \},
\end{equation}
where in particular the $q$-Gaussian distribution is defined as
\begin{math}
p(x,\mu,\sigma)=\exp_{q} (-\frac{(x-\mu)^2}{2 \sigma^2} - \psi(\mu,\sigma) )
\end{math}
(see Appendix, Section \emph{B}, for a connection with the correlation coordinates).

In analogy with the exponential families the $q$-geometry has a dually flat geometrical structure (as its coordinates are orthogonal) and accomplishes the $q$-Pythagorean theorem. The maximiser of the $q$-escort distribution is a Bayesian MAP (maximum a
posteriori probability) estimator as proved in Ref. \cite{Amariqexp}. Moreover, it is possible to generalise
the $q$-structure to any family of probability distributions, because any
parameterised family of probability distributions forms a submanifold embedded in the
entire manifold. Altogether, we can introduce the $q$-geometrical structure to any arbitrary family of probability distributions
and guarantee that the family of all the probability distributions belongs to the $q$-exponential family of distributions
for any $q$ \cite{Amariqexp}. We refer to {Fig~\ref{inf00DD}} for a schematic explanation of the $q$-exponential family.

Next, we take advantage of recent mathematical progress on $q$-geometry (and
$q$-information geometry) to investigate the effect of high - order correlations
on the probability distribution in the asymptotic limit.

\section{Results}
\label{sec:results}

\subsubsection{Beyond pairwise correlations}

In statistical mechanics it is said that we reach the ``thermodynamic limit'' when
the number of particles being considered reaches the limit $N \rightarrow \infty $. The thermodynamic
limit is asymptotically approximated in statistical mechanics
using the so-called central limit theorem (CLT) \cite{sheldon}.
The CLT ensures that the probability distribution
function of any measurable quantity is a normal Gaussian distribution, provided that a
sufficiently large number of \emph{independent random variables}
with exactly the same mean and variance are being considered (see pages 324-330 \cite{sheldon}).
Thus, the CLT does not hold if correlations between random variables cannot be neglected.

Thus, the CLT articulates conditions under which a sufficiently large number of independent and identically
distributed random variables, each with finite mean and variance, can be considered as normally distributed \cite{sheldon}.
In particular, the CLT has been used by Amari and
colleagues \cite{AmariWu} to estimate the joint probability distribution of firing in a neuronal pool
considering the limit of a very large number of neurons. Thus, in their approach pairwise correlations
within the joint distribution of firing are quantified through the covariance $<U_i U_j>$ of
the weighted sum of inputs $U_i$ and $U_j$ of two given pairs of neurons ($i\neq j$, $i=1...N$ and $j=1...N$ ) \cite{AmariWu}.
This is $U_i = \sum_{j=1}^m  w_{ij} -H $, where $w_{ij}$ is the connection weight from the $j^{th}$ input to the $i^{th}$ neuron ($H=E[U_i]$ denotes the mean). Importantly, these $U_i$ are being considered Gaussian due to the CLT \cite{AmariWu}, and thus
the approach quantifies the amount of pairwise correlations through the covariance $<U_i U_j>$.

In the presence of weak or strong correlations of any sort, the CLT has been generalised
in recent publications by M Gell-Mann, C Tsallis, S Umarov, C Vignat, A Plastino
(see:\cite{umarov1,umarov2,vignat,vignat1,vignat2}).
They have proved that when a system with weakly or strongly
correlated random variables is being considered, if we gather a sufficiently large number of such systems together,
the probability distribution will converge
to a $q$-Gaussian distribution \cite{umarov1,umarov2,vignat,vignat1,vignat2}.
This is in agreement with the theorems recently proved by Amari and Ohara \cite{Amariqexp},
which permit the introduction of the $q$-geometrical structure to any arbitrary family of probability distributions,
and guarantee that the family of all the probability distributions belongs to the $q$-exponential family of distributions.

We will use the ``natural extension'' of the central limit theorem (ECLT) proposed in \cite{vignat},
which accounts for cases in which correlations between random
variables are non-negligible. This results in so-called
$q$-Gaussians (instead of Gaussians) as the PDFs in the ECLT,
as proved in Ref.\cite{vignat}:
\begin{equation}
G_{q} (x) =
\begin{cases}
[1 + \frac{ (1-q) (-x^2)}{2}]^{\frac{1}{1-q}} & \text{if  $(1 + \frac{(1-q) (-x^2)}{2}) > 0$,}\\
0 & \text{otherwise}
\end{cases}
\end{equation}
where $q$ is a (problem-dependent) positive real index.
Notice that in the limit of $q=1$ a \emph{normal Gaussian distribution} is recovered as
$\lim_{N -> \infty} (1 + \frac{1}{N})^N = e,$ which can be rephrased as $\lim_{q -> 1} (1 + (1-q))^{\frac{1}{(1-q)}}=e$.
In other words, the CLT is being recovered as $q \rightarrow 1$ \cite{umarov1,umarov2,vignat,vignat1,vignat2}.

Let us now consider the probability of exactly $k=N \cdot r$ (and thus $r=\frac{k}{N}$)
neurons firing within a given time window $\Delta T$ across a population of $N$
neurons. In the framework of the pooled model we have:\\ \begin{math}
P_{r}[r=\frac{k}{N}]= P_{r}\{x_1=x_2=\ldots=x_k=1, x_{k+1}=\ldots=x_N=0\},
\end{math}
where the neuron $ x_i $ is subject to a weighted sum of inputs $u_i$, thus $x_i=1$
if and only if $u_i > 0$ and $x_i=0$ if $u_i \leq 0$. Following
\cite{AmariWu}, the neuronal pool receives common inputs $s_1,
s_2,...s_M$ (as schematised in {Fig~\ref{inf0}}), and $u_i$ is weighted by the common inputs $u_i =
\sum_{j=1}^M w_{ij} s_j - h $, where $w_{ij}$ are randomly assigned connections weights. These $u_i$ are $q$-Gaussian due
of the ECLT \cite{umarov1,umarov2,vignat,vignat1,vignat2}. Considering that the $u_i$ are subject to a
$q$-Gaussian distribution $N_{q}(-h, 1)$, we define in analogy to \cite{AmariWu}
$u_i=\sqrt{1-\alpha} \, v_i +\sqrt{\alpha} \, \varepsilon - h$, for $i=1,..,N$. We take $\alpha=E_{q}[u_i u_j]$ as
a $q$-variance, $h=E_{q}[u_i]$ as the $q$-mean, and two independent $q$-Gaussian random variables
$v_i$ and $\varepsilon$ subject to $N_{q}(0, 1)$ (see \cite{Thistleton} for a detailed description of $q$-Gaussian
random variables).

In the following we will use what is commonly referred to
factorization approach in statistical mechanics \cite{Buyu94,Sandra}, which is applicable in this case as
we are considering weak correlations among neurons and the population of neurons is
homogenous. We name $E_{\varepsilon}$ as the expectation value taken
with respect to a random variable $\varepsilon$, and $P_{r}\{u > 0 | \varepsilon \} $  is the conditional probability for
$\varepsilon$. This allows us to
calculate the probability of having $r=\frac{k}{N}$ neurons firing, separating
the contribution of neurons that are firing
$[P_{r}\{u > 0 | \varepsilon \}]^k$ from those that are silent $[P_{r}\{u \leq 0 | \varepsilon \}]^{N-k} ] $,
as

\begin{eqnarray}
\begin{split}
P_{r}\{x_1=x_2=\ldots=x_k=1,
x_{k+1}=\ldots=x_N=0\}= \\ E_{\varepsilon} [P_{r}\{u_1;u_2;\ldots;u_k > 0,
u_{k+1};\ldots;u_N \leq 0 | \varepsilon \}] \equiv \\ 
                         E_{\varepsilon} [ (\begin{array}{c}
                       N \\
                       k
                     \end{array})[ P_{r}(u > 0 | \varepsilon )]^k [P_{r}(u \leq 0 | \varepsilon )]^{N-k} ]. \\
\label{prprob}
\end{split}
\end{eqnarray}
In order to go beyond the pairwise estimation of \cite{AmariWu} (performed within the CLT framework),
we need to quantify the amount of correlations higher than two in the probability distributions.
If we take the limit of $N \rightarrow \infty$, in the framework of
the $q$-Gaussian ECLT \cite{umarov1,umarov2,vignat,vignat1,vignat2}, instead of the Gaussian CLT as
considered in \cite{AmariWu}, we can define:
\begin{eqnarray}
\label{eqFqq}
{F_{q}(\varepsilon)} \equiv P_r(u>0 | \varepsilon) &=& P_r(u_i >
\frac{h-\sqrt{\alpha} \varepsilon}{\sqrt{1-\alpha}})\\ \nonumber
 &=&\frac{1}{\sqrt{2 \pi}} \int_{\frac{h-\sqrt{\alpha}
    \varepsilon}{\sqrt{1-\alpha}}}^{\infty} \!
\exp_{q}{(-\frac{v^2}{2})} \, dv .
\end{eqnarray}
If we take $q=1$ in the probability of having $r=\frac{k}{N}$ neurons firing in Eq.~(\ref{eqFqq}),
and the distributions within the integral of Eq.~(\ref{eqFqq}) corresponds to normal Gaussian distributions, and we are in
the ``CLT framework''.
On the other hand, the ``ECLT framework'' corresponds to $q > 1$ and in this case the system has weakly or strongly correlated random variables. Thus the distributions within the integral of Eq.~(\ref{eqFqq}) are considered as q-Gaussian distributions, and correlations are quantified through $q$.

Notice that if we consider the limit of the CLT framework ($q=1$), the previous expression reduces to
\begin{eqnarray}
{F_{q=1}(\varepsilon)} = \frac{1}{2} \mathrm{Erfc}(\frac{1}{\sqrt{2}} \frac{h-\sqrt{\alpha} \varepsilon}{\sqrt{1-\alpha}})
\label{erfcc}
\end{eqnarray}
where $\mathrm{Erfc(x)}=\frac{2}{\sqrt{\pi}} \int_x^{\infty} \exp(-t^2) dt$ denotes the complementary error function.
However, if the effect of correlations higher than two is not negligible then according to the ECLT: $q > 1$, thus
\begin{eqnarray}
\exp_{q} (-\frac{v^2}{2}) = \frac{1}{\Gamma(\frac{1}{q-1})} \int_0^{\infty}
dt \, t^{\frac{1}{q-1}-1} \exp({-t - t (q-1) \frac{v^2}{2}}),
\end{eqnarray}
which is known as Hilhorst transform \cite{grads2007}, an integral
representation widely used in generalised statistical
mechanics \cite{Tsallis94}.  Thus $F_{q}(\varepsilon)$ reads
\begin{equation}
\begin{split}
{F_{q}(\varepsilon)} = \frac{1}{\Gamma(\frac{1}{q-1})}
\int_0^{\infty} \int_{\frac{h-\sqrt{\alpha}
    \varepsilon}{\sqrt{1-\alpha}}}^{\infty} dt \, dv \,
t^{\frac{1}{q-1}-1} \exp({-t + t (q-1) \frac{v^2}{2}}).
\label{fvar}
\end{split}
\end{equation}
Using several non-trivial identities between Gauss hypergeometric functions
and the incomplete beta function, we can exactly calculate the integrals expressed above (see Appendix Section \emph{A}, for a detailed description of the math).
And ${F_{q}(\varepsilon)}$ in terms of a beta incomplete function reads
\begin{equation}
\begin{split}
{F_{q}(\varepsilon)}= \frac{1}{{2 \sqrt{2 \pi}}
\sqrt{\frac{q-1}{2}}}  {B_{\frac{1}{1+\xi_0(\varepsilon)}}(\frac{1}{q-1},\frac{1}{2})},
\label{eq:beta}
\end{split}
\end{equation}
where \begin{equation} \xi_0(\varepsilon)=\frac{(q-1)(h-\sqrt{\alpha}\varepsilon)^2}{2({1-\alpha})} \label{eqxi0} .\end{equation}
Eq.(~\ref{eq:beta}) allows us to calculate the probability
(Eq.~\ref{prprob}) of exactly $k$  neurons firing within a given time window $\Delta T$ across a population of $N$
neurons. Notice that the amount of correlations higher than two is quantified through $q >1$, as when $q$ is constrained
to 1 it leads to ${F_{q=1}(\varepsilon)}$, which is reduced to the estimation within the CLT (Eq.~\ref{erfcc}) as in \cite{AmariWu}.

The joint distribution of firing can therefore be estimated as
\begin{eqnarray}
Q_{q}(r) &\simeq& N  P_{r}[r=\frac{k}{N} ] \\&=& N E_{\varepsilon}
\{\left(\begin{array}{c}
                       N \\
                       k
                     \end{array}\right)
[{F_{q}(\varepsilon)} ]^{k} [1- F_{q}(\varepsilon) ]^{N-k} \}.\nonumber
\label{qrhod}
\end{eqnarray}

The expectation value $E_{\varepsilon}$
can be estimated using the saddle point approximation
\cite{Butler07,AmariWu}
\begin{equation}\begin{split} Q_{q}(r) = \sqrt{\frac{1}{ r (1-r)|z_{q}''(\varepsilon_0)|}} \frac{1}{\sqrt{2 \pi}} \exp[N z_{q}(\varepsilon_0)-\frac{{\varepsilon_0}^2}{2} ],
\label{eq:fin}
\end{split}
\end{equation}
where $z_{q}(\varepsilon)=r
\log(\frac{{F_{q}(\varepsilon)}}{r}) + (1-r)
\log(\frac{1-{F_{q}(\varepsilon)}_{q} }{1-r})$. Within the
saddle point approximation: $\varepsilon_0 = \mathrm{\arg\max_{\varepsilon\epsilon\mathbb{R}} }\left[
z_{q}(\varepsilon)\right]$ and
$\frac{dz(\varepsilon)}{d\varepsilon}=0$. The solution is $\varepsilon_0 =
{F_{q}^{-1}(r)},$ which implies
$r={F_{q}(\varepsilon_0)},$ where $r$ goes between $[0,1]$ and
$\varepsilon_0$ is defined for all real numbers.
$\xi_0(\varepsilon)$ (Eq.~\ref{eqxi0}) can take different values with
$\alpha$ and $h$. Additionally, $\varepsilon_0 = {F_{q}^{-1}(r)}$ depends
on the degree of correlation of the network architecture, which is quantified by $q$.
Fig~\ref{inf1a} shows the behaviour of $ Q_{q}(r)$
for $q=1.3$ in comparison to $q=1$. Notice that a higher degree of correlation, $q > 1 $,
corresponds to a more widespread distribution. This is in agreement with the idea of Amari and Nakahara \cite{AmariWu}
that when taking a very large number of neurons higher-order correlations are needed to reproduce the behaviour of a
widespread distribution.

Estimating Eq.~(\ref{eq:fin}) for $q > 1$ requires a non trivial approach: finding $\varepsilon_0 = {F_{q}^{-1}(r)}$, and
from Eq.~(\ref{eq:beta}) this means to estimate the inverse incomplete beta function of
${B_{\frac{1}{1+\xi_0(\varepsilon)}}(\frac{1}{q-1},\frac{1}{2})}$ (multiplied by the factor
$\frac{1}{{2 \sqrt{2 \pi}}\sqrt{\frac{q-1}{2}}}$). The inverse beta functions are tabulated in exhaustive detail,
thus we can perform the numerical estimation of Eq.~(\ref{eq:fin}). It is important to note that the incomplete
 beta function subroutines in matlab are normalised as
\begin{math}
{I_{w}(p,s)}= \frac{\int_{0}^{w} t^{s-1} (1-t)^{s-1} dt}{B_{1}(p,s)},
\end{math}
in which
\begin{math}
{B_{1}(p,s)} = \frac{\Gamma(p)\Gamma(s)}{ \Gamma(p+s)}
\end{math}
and therefore we should also multiply by ${B_{1}(p,s)}$ to estimate Eq.~(\ref{eq:beta}) numerically.
Notice from Eq.~(\ref{erfcc}) that if we take $q=1$, then $F_{q=1}^{-1}(r)$ is
reduced to the estimation of an inverse complementary error function that is also a non-trivial mathematical
operation. When $q=1$ we are in the CLT framework and just estimating pairwise correlations,
and we are potentially missing higher-order correlations. In contrast, ${F_{q}^{-1}(r)}$ allows us to quantify
the amount of higher-order correlations as we are within the ECLT
framework.
It is important to point out that when $q=1$, our previous findings reduce to the
CLT limit estimations of Amari \cite{AmariWu}.  Thus, for real
data, one can test for the presence of higher-order correlations by
measuring the distribution of activity in multiunit recordings, and
fitting  $q$, which represents the amount of higher-order
correlations present in the distribution of firing. One can show by
simple comparison how statistically different from the $q=1$ case
the measured distribution is.

In the next section (Experimental Results) we test the applicability of Eq.~(\ref{eq:fin}) by measuring the spiking rate
of multiunit activity in all non-overlapping windows of length $\Delta T$.
We fit the parameter $q$ to find the best-fitting function $Q_q(r)$ in Eq.~(\ref{eq:fin})
for the experimental distribution. We then test the hypothesis of absence of higher-order correlation by comparison
with a fit with $q$ constrained to equal 1.

In Section~\ref{sec:simulation} we consider a network simulation model in which the
number of interconnected neurons is a parameter under
control. We then evaluate the hypothesis that Eq.~(\ref{eq:fin}) permits
characterising the internal dynamics of the network for
a spatio-temporal neuronal data set and quantifying
the degree of higher-order correlations through $q$.

A measure that is also particularly interesting in this context, since it
was used in Ref. \cite{linnik} to give an information theoretic proof of
the CLT, is Fisher information
\begin{eqnarray}
 I(Q_{q}) = \int_0^{1} dr \frac{[\frac{\partial Q_{q}(r)}{\partial r}]^2}{Q_{q}(r)}.
 \label{fish}
\end{eqnarray}
This measure is useful for detecting dynamical changes in the PDFs
(i.e. a sharper probability distribution function would tend to
have higher Fisher information than a more widespread PDF). Discretising Eq.~(\ref{fish}) as in Ref.
\cite{ferri2009}, using a grid of size $M$ to calculate the distribution, one obtains
\begin{eqnarray}
 I(Q_{q}) = \frac{1}{4} \sum_{i=1}^{M-1} 2 \frac{({Q_{q}}_{i+1} - {Q_{q}}_{i})^2}{({Q_{q}}_{i+1} + {Q_{q}}_{i})}.
 \label{fish2}
\end{eqnarray}

This provides us with an information theoretic measure to quantify the dynamical changes of the distribution.
It is usually accepted when considering Fisher information and population codes of independent neurons that
they encode information through bell-shaped tuning curves, and that the mean firing rate of a neuron is a Gaussian
function of some variable (i.e the external stimuli). In this case the slope of the distribution increases as
the width of the tuning curve becomes smaller, and then a shaper distribution would have a higher slope
and thus higher Fisher information. However, this would the case if noise correlations are independent across
neurons and independent of the tuning width, as sharpening in a realistic network does not guarantee
a higher amount of Fisher information \cite{peggy}.

\subsection{Experimental results}
\label{sec:ExperimentalResults}

To provide an initial test of the applicability of this approach to the characterisation of higher-order correlations in pools
of activity obtained from real neurophysiological data, we used a silicon microfabricated linear electrode array
(NeuroNexus Inc., Michigan, USA) to record multiunit activity from the mouse barrel cortex (see Materials and methods).
We used 5 minutes of spontaneous spiking cortical activity of three adult mice, in which 16 electrodes were placed to record
the multiunit neural activity (three data sets with 16 different channels each). In the following, we apply our formalism to provide
a statistical characterisation of higher-order correlations in pools of multiunit neural activity for each of the 16 different
channels of the three data sets (thus in total 48 different recordings of 5 minutes in length), where we considered four
non-overlapping window lengths of $\Delta T= 25$ ms, $\Delta T= 50$ ms, $\Delta T=100$ ms, and $\Delta T=200$ ms.
Our selection of these non-overlapping windows is related to the time windows typically
used to investigate spike correlations and the firing rate distributions in Ref. \cite{Panzeri99,panzeri01visu,Treves}.
The time windows of $\Delta T=25$ ms and $\Delta T=50$ ms are similar values to those used to investigate
spike correlations in Ref. \cite{Panzeri99,panzeri01visu}, and broader time windows of
$\Delta T=100$ ms and $\Delta T=200$ ms have also been used to investigate the firing rate distributions
\cite{Treves}, and functional role of spike correlations \cite{panzeri01visu}. The window length
of $\Delta T = 25$ ms is close to the values used in \cite{Shlens06,Schneidman_Nature_06} to investigate the effect of
spike correlations.

We estimated the normalised firing distribution $Q_{q} (r)$, for experimental data, by measuring the spiking rate in all
non-overlapping windows of length $\Delta T$ (here 25, 50, 100, and 200 ms). We then fitted the parameter $q$, representing
the extent of higher-order correlations present in the neuronal pool, to find the best-fitting function $Q_q(r)$
(as in Eq.~(\ref{eq:fin})). Values of $q$ equal to unity imply the absence of higher-order correlations in the system.
Fig~\ref{inf2} shows the fitted ${q}$ extracted for 3 different sets of 16 simultaneously recorded channels of multiunit
activity, of increasing depth in the cortex (in 50 $\mu$m steps). For each set, the recordings were taken using different
time windows $\Delta T=25 $ ms (panels \emph{a-c}), $\Delta T=50 $ ms (panels \emph{d-f}), $\Delta T=100 $ ms (panels \emph{g-i}) and $\Delta T=200 $ ms (\emph{j-l}). Notice that the maximum and minimum values of $q$ corresponds to channel 15 in panel \emph{i} and channel 12 in panel \emph{l},
respectively, for time windows of $\Delta T=100 $ ms and $\Delta T=200 $ ms.

In order to understand how well the model proposed in Eq.~(\ref{eq:fin}) fits the experimental data, we compared it with
a fit with $q$ constrained to equal 1. We define the normalised firing rate as $r \equiv \frac{\langle k \rangle}{N_{max}}  $,
where $\langle k \rangle$ is the mean firing rate and $N_{max}$ is the maximum number of spikes. Figs~\ref{inf3} and ~\ref{inf4}
show the experimental spontaneous distribution of firing $Q_{q}(r)$ as a function of $r$ (on logarithmic scale) for those
channels with the maximum and the minimum values of $q$, respectively. That is, we use the estimated $q$ and the
parameters $\alpha$ and $h$ that give the best fit for the distribution.
The optimisation fitting criterion is the normalised mean squared error (NMSE), and the default error value is lower
than 0.05 ($p-value < 0.05$). It is apparent that for all 25, 50, 100 and 200 ms time windows
(Figs~\ref{inf3}-\ref{inf4} \emph{a},\emph{b},\emph{c} and \emph{d}, respectively), the
theoretical curve is a remarkably good fit to the experimental distribution ($p-value < 0.05$ ). In contrast, the $q=1$
curve does not come close to modelling the data satisfactorily over a wide range of firing rates.

To detect the dynamical changes in the probability distributions, we estimate Fisher information as in Eq.(\ref{fish2}).
On the one hand Fig~\ref{inf5} shows Fisher Information versus $q$ for the entire data set.
Fig~\ref{inf5a}{\emph{a}} shows Fisher information averaged
over the 48 different channels, which becomes higher for smaller time windows.
Notice from Fig~\ref{inf5a}{\emph{b}} that when considering the case $q=1$,
Fisher information takes higher values for 50, 100 and 200 ms time windows than for those in which $q >1$
is considered (Fig~\ref{inf5a}{\emph{a}}).
This is not the case for 25 ms time windows in which Fisher information takes a much higher value for
$q >1$ than for those with $q$ constrained to be equal to 1.
As expected, Fisher information is more substantial at shorter time windows, where the fine temporal precision at which the
spikes may synchronise has a significant effect.

Overall our findings show that spontaneous distributions of firing with $q > 1$ should be considered to
accurately reproduce the experimental data set. The main advantage of this method is that through the inclusion of a deformation
parameter $q$, which accounts for correlations higher than two within the probability distributions, we can quantify
the degree of correlation and reproduce the experimental behaviour of the distribution of firing.

If we considered single neuron trial to trial fluctuations for a fixed stimulus in spike count fluctuations, the extent
of noise correlations would depend on the width of the tuning curves as correlations come up from common
inputs \cite{peggy}. Thus, in this case Fisher information would depend on the sharpening of the distribution and
also on the noise correlation, and therefore a sharper distribution would not necessarily imply a higher
amount of Fisher information.
This effect is shown, for instance, in Fig~\ref{inf4}{\emph{d}} ($\Delta T=200$ ms) and
Fig~\ref{inf4}{\emph{b}} ($\Delta T=50$ ms); notice that \emph{d} corresponds to a value of $q$ smaller than \emph{b}. The
firing distribution of Fig~\ref{inf4}\emph{d} has a higher slope than the one presented in Fig~\ref{inf4}\emph{b}, but this higher
slope does not have a correspondence with a higher value of Fisher information (Fisher info equal to 0.0150 bits
in Fig~\ref{inf4}\emph{d}, and Fisher Info equal to 0.0395 bits in Fig~\ref{inf4}\emph{b}).

Summing up, we presented a formalism that provides us with an estimate of the degree of correlation for the distribution of
firing within pools of multiunit neural activity. This method allows us to naturally distinguish how far this distribution of
firing is from the one we would obtain if each neuron contributes within CLT ($q=1$). It permits us, therefore,
to quantify the amount of correlation significantly reducing the number of parameters associated with the correlation coordinates.

\subsection{A Network simulation model}
\label{sec:simulation}

To test further our theoretical approach, we apply our formalism
to a network simulation model in which the number for interconnected neurons is a parameter under
control.
We then analyse a simple network model in which neurons receive common overlapping inputs as in Fig~\ref{inf0},
considering that each neuron can be interconnected randomly with more than two neurons.
This will allow us to also test the hypothesis of Amari and collaborators that
weak higher-order interactions of almost all orders are required to
realise a widespread activity distribution of a large population of neurons \cite{AmariWu}.

The network simulation we use is the one developed in \cite{Izhik06}, which consists of cortical spiking neurons
with axonal conduction delays and spike timing-dependent plasticity (STDP). Each neuron in the network is described by the
simple spiking model \cite{Izhik03}

\begin{equation}
v' =0.04 v^2 + 5v + 140 . u + I
\end{equation}
\begin{equation}
u' =a(bv-u)
\end{equation}
with the auxiliary after-spike resetting
\begin{equation}
if v \geq +30 mV, then
\begin{cases}
v \leftarrow c \\
u \leftarrow u + d. \\
\end{cases}
\label{izhi}
\end{equation}
Where $v$ is the membrane potential of the neuron, and $u$
is a membrane recovery variable, which accounts for the activation
of $K+$ ionic currents and inactivation of $Na+$ ionic currents, and gives
negative feedback to $v$. After the spike reaches its apex at +30 mV (not to be confused with the firing threshold)
the membrane voltage and the recovery variable are reset according to Eq.~(\ref{izhi}). The variable $I$ accounts for
the inputs to the neurons \cite{Izhik03,Izhik06}.

Since we cannot simulate an infinite-dimensional system on a finite-dimensional lattice, we choose a network with
a finite-dimensional approximation taking a time resolution of 1 ms.
The network consists of $N = 1000$ neurons with the first
$Ne = 650$ of excitatory RS type, and the remaining $Ni=350$ of inhibitory FS
type \cite{Izhik03}. Each excitatory neuron is connected to $M=2$,$3$,$20$,$40$,$60$,$80$,$100$,$120$,$140$ and $160$
random neurons, so that the probability of connection is $M/N = 0.002$, $0.003$,$0.02$,$0.04$,$0.06$,$0.08$,$0.1$,$0.12$,$0.14$ and $0.16$.
Each inhibitory neuron is connected to $M=2$,$3$,$20$,$40$,$60$,$80$,$100$,$120$,$140$ and $160$ excitatory
neurons only. Synaptic connections among neurons have fixed conduction
delays, which are random integers between 1 ms and 20 ms.

The main idea of this simulation is to investigate Amari's hypothesis that correlations of almost all orders are
needed to realise the widespread distribution of firing when a large number of neurons is considered and
to show the behaviour of the parameter $q$ when the number of interconnected neurons is changed.

In order to show that the probability distribution in the thermodynamic limit is not
realised even when pairwise interactions, or third-order interactions, exist, we estimate the
joint probability distribution of firing when each excitatory/inhibitory neuron is interconnected to M=2 and 3 neurons.
We run 10 minutes of simulated spiking activity, considering a window length of $\Delta T = 25$ ms, which is a time window
close to the one used by \cite{Panzeri99,panzeri01visu,Shlens06,Schneidman_Nature_06} to investigate the effect of correlations.
As shown in Fig~\ref{inf00Da}\emph{a} $q=1$ is a remarkably good fit to the simulated distribution of firing ($p-value < 0.05$ )
when each excitatory/inhibitory neuron is connected to $M=2,3$ random neurons. Notice that we find
no difference in the probability distribution of firing when M=2 and M=3 are considered.
However, when the number of interconnected neurons becomes higher, the distribution of firing becomes
more widespread (see Fig~\ref{inf00Da}\emph{b}) and $q$ increases as the number of interconnected neurons
increases (Fig~\ref{inf00Db}). In the current simulation we took 1000 neurons, however,
when we chose 100-200 neurons, we were also able to capture the
effect of the higher-order correlation if the parameter M was large enough to produce
a widespread distribution (see Appendix, Section \emph{C}).

\section{Discussion and conclusions}
\label{discusion}

Approaches using binary maximum entropy models at a pairwise level have been developed considering
a very large number of neurons on short time scales \cite{Schneidman_Nature_06,Shlens06,Shlens09}. These models can capture
essential structures of the neural population activity, however, due to their pairwise nature their generality
has been subject to debate \cite{Bethge08,roudi09,victor10}. In particular, using an information geometrical approach,
E. Ohiorhenuan and J. D. Victor have shown the importance of the triplets
to characterise scale dependence in cortical networks. They introduced a measure
called ``strain'' that quantifies how a pairwise-only model must be ``forced'' to accommodate the observed triplet firing patterns
\cite{victor10}. Thus, although models accounting for pairwise interactions have proved able to capture some of the most important features of  population
activity at the level of the retina \cite{Schneidman_Nature_06,Shlens06}, pairwise models are not enough to
provide reliable descriptions of neural systems in general
\cite{Bethge08,roudi09,victor10,victorN10,schneidman11}.

Very little is known about how the information saturates as the number of neurons increases.
It has been pointed out by Amari and colleagues \cite{AmariWu} that as the number of neurons increases, pairwise or
triplet-wise correlations are not enough to realise a widespread distribution of firing.
Understanding how neural information saturates as the number of neurons increases would require the
development of an appropriate mathematical framework to account for correlations higher than two
in the thermodynamic limit.

It is important, therefore, to develop an appropriate mathematical
approach to investigate systems with a large number of neurons, which could account for correlations of
almost all orders within the distribution of firing.

In this paper we present a theoretical approach to quantify the extent of higher than pairwise spike correlation in pools of multiunit activity when taking the limit of a very large number of neurons. In order to do this,
we take advantage of recent mathematical progress on $q$-geometry to investigate, in the asymptotic limit, the effect of higher-order
correlations
on the probability distributions within the ECLT framework \cite{Amariqexp,umarov1,umarov2,vignat,vignat1,vignat2}. The main basis of our formalism is that when taking the limit of a very
large number of neurons within the framework of the CLT as in \cite{AmariWu}, we are losing information
about higher-order correlations. Thus, in the new theoretical approach
we take the limit of a very large number of neurons within the framework of the ECLT,
instead of the CLT. The inclusion of a deformation parameter $q$ in the ECLT framework allows us to reproduce
remarkably well the experimental distribution of firing and to avoid the sampling size problem of Eq.~(\ref{eq:pthetafull})
due to the exponentially increasing number of parameters.

We estimated the normalised firing distribution $Q_{q} (r)$, from multiunit recordings, and
fitted the parameter $q$ to find the best-fitting function $Q_q(r)$ (as in Eq.~(\ref{eq:fin})). We showed by
simple comparison how statistically different from the $q=1$ case the measured distribution is
(as it is reduced to the CLT pairwise estimations of \cite{AmariWu}).
Our theoretical predictions provided a remarkably good fit for the experimental distribution. We showed that
the parameter $q > 1$ can capture higher-order correlations, which are salient features of the distribution of firing,
when applied to our multiunit recording data obtained from mouse barrel cortex. As higher-order correlations were present
in the data, the
distributions in the CLT framework do not fit the experimental data well.

Staude and collaborators \cite{staude2007,staude2010a,staude2010b,staude2011} have introduced a quite powerful approach based on continuous-time point process to investigate higher-order correlations in non-Poissonian spike trains.
Higher-order interactions are very important to investigate the neuronal interdependence in the cortex and
at population level of the neuronal avalanches \cite{plenz}. More specifically, Plenz and collaborators \cite{plenz} have developed a
very powerful theoretical framework based on a Gaussian interaction model that takes into account the pairwise correlations
and event rates and by applying a intrinsic thresholding permit distinguishing higher order interactions.
In this approach the pattern probabilities for the so-called ``DG model'' were estimated using the cumulative
distribution of multivariate Gaussians and showed a high fitting precision of the experimental data.
Our current theoretical formalism relies on a different basis: the recent progress made on the ECTL, and the main goal of our approach, is to provide a quantification of the amount of correlations higher than two
when considering a large population of neurons. Thus using mathematical tools of non extensive statical mechanics
\cite{Tsallis94,vignat,vignat1,vignat2,AmariMono,TsallisGell-Mann,AmariWu,Stumpf,umarov1,umarov2}, we introduced an approach that provides a quantification of the degree of
higher order correlation for a very large number of neurons.

It would be interesting to develop
in the close future a paper showing a careful comparison of our current method with the one developed by Plenz and collaborators. This would
help to investigate the possible link of the quantification through the $q$ parameter with the ``DG model''\cite{plenz} and to
gain more insights for future analysis.
Evaluating the degree of $q$ can help to understand
further the processing of information in the cortical network and to get more understanding of the non-linearities of
information transmission within a neuronal ensemble.
Although formally speaking the multiunit recordings are not in the``thermodynamic limit'', the current methodology presented in this paper is a completely new theoretical approach for theoretical neuroscience. And as more data become available for a very large number of neurons, i.e. including evoked activity to sensory stimuli, our theoretical approach could provide an important mathematical tool to evaluate important questions such as how
quickly neural information saturates as the number of neurons increases,
and whether it saturates at a level much lower than the amount of information available in the input.

The pooling process can be taken to reflect the
behaviour of a simple integrate-and-fire neuron model in reading out
the activity of an ensemble of neurons, or alternately the recording of multiunit
neural activity, without spike sorting. As we cannot know what the exact
number of neurons in our multiunit recordings is, we approximated this number
by the maximum number of spikes. In the pooling approach we developed within the ECLT
framework we assumed the asymptotic limit of a very large number of neurons.
Our method was developed within the information geometry framework, similar to that described in \cite{victor10}. However,
it is important to remark three important differences between the method developed in \cite{victor10}
and our current theoretical approach: first, we assumed homogeneity across neurons; second, our approach
was developed assuming a very large number of neurons within the ECLT framework, and third, our approach accounted for cases
in which a widespread probability distribution is not realised even when pairwise interactions, or third-order interactions, exist.
The significance of our approach is that it allows us to extend
analytically solvable models of the effect of correlations higher than two and to compute their scaling properties
in the case of a very large number of neurons. Our treatment provides a quantification of the degree of correlation
within the probability distribution, which is summarised in a single $q$ parameter, thus avoiding the sample
 size problem that constitutes the main intractability
reason of the approaches presented in  Eq.~(\ref{eq:pthetafull}).

To contrast our current data analysis with a case in which the network structure is known a priori,
we tested our theoretical approach using a network simulation model in which the number for interconnected neurons
is a parameter under control. We then analysed a simple network model in which neurons receive common overlapping inputs
and considering that each neuron can be interconnected randomly with more than two neurons.
We showed that in the specific network model, high connectivity is required to get a widespread distribution,
which is in agreement with the hypothesis of Amari \cite{AmariWu} that weak higher-order interactions of almost all orders are required for
realising a widespread activity distribution in the ``thermodynamic limit'' \cite{AmariWu}.
Moreover, our results are in agreement with the hypothesis of Amari and colleagues \cite{AmariWu}
that the widespread probability distribution in the thermodynamic limit is not
realised even when pairwise interactions, or third-order interactions, exist. Correlations of almost
all orders are then needed to realise the widespread activity distribution of a very large population
of neurons. In our current simulation we took 1000 neurons, which may be considered
quite a large number in comparison to the number of neurons that a multiunit recording might capture.
However, when we chose 100-200 neurons in our simulation, we were also able to capture the effect of the higher-order
correlation if the number of interconnected neurons was large enough to produce a widespread
distribution (see Appendix, section \emph{C}). Thus, as higher-order correlations
were present in the simulated data set, a very weird network architecture will be required to force $q=1$,
and thus in this case the distribution of firing in the CTL framework does not come even close to fitting the simulated data.

The model we developed using an information - geometric approach within the ECLT framework, and together with Fisher information
estimations, in principle could allow population codes involving higher-order correlations to be studied in the thermodynamic limit,
in the same way as other authors have done for the second order case \cite{dayan09,peggy,averbeck06}. This is particularly
interesting
as in most cases pairwise models do not provide reliable descriptions of true biological systems \cite{roudi09}.
Thus our approach
could be of help to gain further insights into the role of high-order correlations in information transmission
for very large systems, and could also be an important mathematical tool to evaluate whether the evoked activity may induce
plasticity effects on the network when compared to the spontaneous signal.
Applying our formalism to a data set obtained from mouse barrel cortex using multiunit
recordings, we showed that a simple estimation of the deformation parameter $q$
attached to the probability distribution of firing can answer us how significant the degree of higher-order spike correlations
is within pools of neural activity.
In our current analysis we show that higher - order correlations are prevalent but they do not, in general, improve the
accuracy of the population code when considering a large number of neurons. Overall our findings show
that Fisher information increases as the time window decreases, which would involve an easier discrimination
task when the time windows become shorter.

Summarising, we presented an information-geometric approach to quantify the degree of spike
correlations higher than two in pools of multiunit neural activity. Our proposed formalism provides a statistical
characterisation of the amount of high-order correlations through the $q$ parameter, avoiding the sampling problem
of the pooling approach when high-order correlations in the thermodynamic limit are considered.

\section{Materials and methods}

Recordings were made from adult female C56BL/6 mice, of 2-3 months of age. The animals were maintained in the Imperial College animal facility and used in accordance with UK Home Office guidelines. The experiments were approved by the UK Home Office, under Project License 70/6516 to S. R. Schultz. Mice were sedated by an initial intraperitoneal injection of urethane (1.1 g/kg, 10 \% w/v in saline) followed by an intraperitoneal injection of 1.5 ml/kg of Hypnorm/Hypnovel (a mix of Hypnorm in distilled water 1:1 and Hypnovel in distilled water 1:1; the resulting concentration being Hypnorm:Hypnovel:distilled water at 1:1:2 by volume), 20 minutes later. Atropine (1 ml/kg, 10 \% in distilled water) was injected subcutaneously. Further supplements of Hypnorm (1 ml/kg, 10 \% in distilled water) were administered intraperitoneally if required. Their body temperature was maintained at $37\pm0.5 ^{\circ}$C  with a heating pad. A tracheotomy was performed and an endotracheal tube (Hallowell EMC) was inserted to maintain a clear airway as previously described \cite{Moldestad2009}. After the animal was fixed on stereotaxic frame, a craniotomy was performed, aimed at above barrel C2. A small window on the dura was exposed to allow insertion of the multi-electrode array. The exposed cortical surface was covered with artificial cerebrospinal fluid (in mM: 150 NaCl, 2.5 KCl, 10 HEPES, 2 CaCl2, 1 MgCl2; pH 7.3 adjusted with NaOH) to prevent drying. The linear probe (model: A1x16-3mm-50-413, NeuroNexus Technologies) was lowered into the brain perpendicularly to the cortical surface, until all 16 electrode sites were indicating multiunit neural activity, and allowed to settle for 30 minutes before recording began.

\section*{Acknowledgments}
Research supported by PIP 0255/11 (CONICET), PIP 1177/09 (CONICET) and Pict 2007-806 (ANPCyT), Argentina (FM-MP), BBSRC DTA studentship (EP) and EPSRC EP/E002331/1, UK (SRS).

\appendix

\section*{Appendix}
\setcounter{equation}{0}
\renewcommand{\theequation}{A-\arabic{equation}}

\subsection*{A. Estimation of ${F_{q}(\varepsilon)}$ and $Q_{q}(r)$}

We take the limit of $N \rightarrow \infty$, in the ``ECLT framework''
\cite{umarov1,umarov2,vignat,vignat1,vignat2}, instead of the Gaussian CLT as
considered in \cite{AmariWu}, we can define:
\begin{eqnarray}
\label{fvarxxx1}
{F_{q}(\varepsilon)} \equiv P_r(u>0 | \varepsilon) &=& P_r(u_i >
\frac{h-\sqrt{\alpha} \varepsilon}{\sqrt{1-\alpha}})\\ \nonumber
 &=&\frac{1}{\sqrt{2 \pi}} \int_{\frac{h-\sqrt{\alpha}
    \varepsilon}{\sqrt{1-\alpha}}}^{\infty} \!
\exp_{q}{(-\frac{v^2}{2})} \, dv
\end{eqnarray}
Notice that if we consider the limit of the CLT ($q=1$), the previous expression reduces to
\begin{eqnarray}
{F_{q=1}(\varepsilon)} = \frac{1}{2} \mathrm{Erfc}(\frac{1}{\sqrt{2}} \frac{h-\sqrt{\alpha} \varepsilon}{\sqrt{1-\alpha}}).
\end{eqnarray}
where $\mathrm{Erfc(x)}=\frac{2}{\sqrt{\pi}} \int_x^{\infty} \exp(-t^2) dt$ denotes the complementary error function.
However, if the effect of higher order correlations than two are not negligible then according to the ECTL
 $q > 1$, thus
\begin{equation}
\exp_{q} (-\frac{v^2}{2}) = \frac{1}{\Gamma(\frac{1}{q-1})} \int_0^{\infty}
dt \, t^{\frac{1}{q-1}-1} \exp({-t - t (q-1) \frac{v^2}{2}})
\end{equation}
which is known as Hilhorst transform \cite{grads2007}, an integral
representation widely used in generalized statistical
mechanics \cite{Tsallis94}.  Thus $F_{q}(\varepsilon)$ reads
\begin{equation}
\begin{split}
{F_{q}(\varepsilon)} = \frac{1}{\Gamma(\frac{1}{q-1})}
\int_0^{\infty} \int_{\frac{h-\sqrt{\alpha}
    \varepsilon}{\sqrt{1-\alpha}}}^{\infty} dt \, dv \,
t^{\frac{1}{q-1}-1} \exp({-t + t (q-1) \frac{v^2}{2}})
\end{split}
\end{equation}
and substituting $w={{\sqrt{\frac{t(q-1)}{2}}}} v$, we can rewrite

\begin{equation}
\begin{split}
\int_{\frac{h-\sqrt{\alpha} \varepsilon}{\sqrt{1-\alpha}}}^{\infty} dv
 \exp\left({-t - t (q-1) \frac{v^2}{2}}\right) =
\frac{\exp\left({-t }\right)}{\sqrt{\frac{t(q-1)}{2}}}
\frac{\sqrt{\pi}}{2}\left(1- \mathrm{Erf}\left[\sqrt{\frac{t(q-1)}{2}}
\left(\frac{h-\sqrt{\alpha}\varepsilon}{\sqrt{1-\alpha}}\right)\right] \right)
\end{split}
\end{equation}
where $\mathrm{Erf(x)}$ denotes the error function,
\begin{eqnarray}
\mathrm{Erf}(x) &=& 1 - \frac{\Gamma(\frac{1}{2},x^2)}{\sqrt{\pi}} \nonumber \\
 &=& \frac{2}{\sqrt{\pi}} \int_0^x \exp(-t^2) dt
\end{eqnarray}
and \begin{eqnarray} \Gamma(\frac{1}{2},x^2) = 2 \int_x^{\infty}
  \exp({- t^2}) dt. \end{eqnarray} Thus, using
\begin{eqnarray}
\begin{split}
\int_0^{\infty} x^{\mu-1} \exp({-\beta x}) \, \Gamma(\nu,\alpha x) dx =
\nonumber {\frac{\alpha^{\nu}  \Gamma(\mu + \nu)}{\mu (\alpha +
    \beta)^{\mu + \nu} }} \,
          {_2F_1}(1,\mu+\nu,\mu+1;\frac{\beta}{\alpha+\beta})
\end{split}
\end{eqnarray}
in Eq.~(\ref{fvarxxx1}) where $Re(\alpha+\beta)>0$,$Re(\mu)>0$, $Re(\mu+\nu)>0$ and $_2F_1$
denotes the Gauss hypergeometric function, we can derive a compact
expression for $F_{q}$ as
\begin{eqnarray}
\begin{split}
{F_{q}(\varepsilon)}= \zeta(\varepsilon)
\, {_2F_1(1,\frac{1}{q-1}; \frac{1}{q-1} +\frac{1}{2} ;
  \frac{1}{1+\xi_0(\varepsilon)})}
\end{split}
\label{hipppox}
\end{eqnarray}
where we named
\begin{eqnarray}
\zeta(\varepsilon) = \frac{\sqrt{\xi_0(\varepsilon)}}{{2 \sqrt{2 \pi}}
  (\frac{1}{q-1}-\frac{1}{2})\sqrt{\frac{q-1}{2}}(1 +
  \xi_0(\varepsilon))^\frac{1}{q-1} }
\end{eqnarray}
and
\begin{eqnarray} \xi_0(\varepsilon)=\frac{(q-1)(h-\sqrt{\alpha}\varepsilon)^2}{2({1-\alpha})}. 
\end{eqnarray}

Making use of the fact that the hypergeometric functions accomplish the following identity \begin{equation} {_2F_1(a,b;c ; z)}={_2F_1(b,a;c ; z)}=(1-z)^{c-a-b}  {_2F_1(c-a,c-b;c ; z)} \end{equation} (see \cite{erdelyi}), we can name: $a=1$; $b=\frac{1}{q-1}$;$c=\frac{1}{q-1}+\frac{1}{2}$;$z=\frac{1}{1+\xi_0(\varepsilon)}$ and therefore
 \begin{equation} {_2F_1(1,\frac{1}{q-1};\frac{1}{q-1}+\frac{1}{2} ;\frac{1}{1+\xi_0(\varepsilon)} )}=( \frac{\xi_0(\varepsilon)}{1+\xi_0(\varepsilon)})^{-\frac{1}{2}} {_2F_1(\frac{1}{q-1}-\frac{1}{2},\frac{1}{2};\frac{1}{q-1}+\frac{1}{2} ; \frac{1}{1+\xi_0(\varepsilon)})} .\end{equation}
Then Eq.~(\ref{hipppox}) reads as,
\begin{equation}
\begin{split}
{F_{q}(\varepsilon)}= \frac{1}{{2 \sqrt{2 \pi}}
  (\frac{1}{q-1}-\frac{1}{2})\sqrt{\frac{q-1}{2}}(1 +
  \xi_0(\varepsilon))^{\frac{1}{q-1}-\frac{1}{2}} } \,
{_2F_1(\frac{1}{q-1}-\frac{1}{2},\frac{1}{2};\frac{1}{q-1}+\frac{1}{2} ; \frac{1}{1+\xi_0(\varepsilon)})}.
\end{split}
\end{equation}

Notice that the incomplete beta function \cite{erdelyi} is defined as
\begin{equation}
\begin{split}
{B_{w}(p,q)}= \int_{0}^{w} t^{p-1} (1-t)^{q-1} dt
            = \frac{1}{p} w^p
\, {_2F_1(p , 1-q ; p+1, w)}
\end{split}
\end{equation}
and therefore naming $p=\frac{1}{q-1}-\frac{1}{2}$,$q=\frac{1}{2}$ and $w=\frac{1}{1+\xi_0(\varepsilon)}$ we can
rewrite
\begin{equation}
\begin{split}
{B_{\frac{1}{1+\xi_0(\varepsilon)}}(\frac{1}{q-1},\frac{1}{2})}= \frac{1}{(\frac{1}{q-1}-\frac{1}{2})(1 +
  \xi_0(\varepsilon))^{\frac{1}{q-1}-\frac{1}{2}} } \,
{_2F_1(\frac{1}{q-1}-\frac{1}{2},\frac{1}{2};\frac{1}{q-1}+\frac{1}{2} ; \frac{1}{1+\xi_0(\varepsilon)})}.
\end{split}
\end{equation}
This allows us to rewrite ${F_{q}(\varepsilon)}$ in terms of a beta incomplete function with dependence on $q$, as
\begin{equation}
\begin{split}
{F_{q}(\varepsilon)}= \frac{1}{{2 \sqrt{2 \pi}}
\sqrt{\frac{q-1}{2}}}  {B_{\frac{1}{1+\xi_0(\varepsilon)}}(\frac{1}{q-1},\frac{1}{2})}.
\end{split}
\end{equation}

The distribution of firing is approached
as

\begin{eqnarray}
Q_{q}(r) &\simeq& N  P_{r}\{r=\frac{k}{N} \} \\&=& N E_{\varepsilon}
\{\left(\begin{array}{c}
                       N \\
                       k
                     \end{array}\right)
[{F_{q}(\varepsilon)} ]^{k} [1- F_{q}(\varepsilon) ]^{N-k} \}.\nonumber
\end{eqnarray}

Using that, {$ \left(\begin{array}{c} N \\ k
                     \end{array}\right) \cong \frac{\exp{(- N  r  \log(r) - N  (1-r)  \log(1-r) )}}{\sqrt{2 \pi N  r (1-r) }}$ }
in the limit of large $N$ we can write \begin{equation}\begin{split} Q_{q}(r) =
    \sqrt{\frac{N}{(2\pi)^2 r(1-r)}} \int_{-\infty}^{\infty}
    d\varepsilon \\ \exp\{ {N [r
        \log(\frac{{F_{q}(\varepsilon)}}{r}) + (1-r)
      \log(\frac{1-{F_{q}(\varepsilon)} }{1-r}) ]-
    \frac{{\varepsilon}^2}{2}}\}
\label{eq:bef}
\end{split}
\end{equation}

But notice that, in the definition of $Q_{q}(r)$ the standard
exponential is used since the correlation effects were previously
included through the deformation parameter $q$ within
${F_{q}(\varepsilon)}$ (where we have used the ECLT \cite{vignat}). The previous integral (Eq.~\ref{eq:bef})
is solved using the saddle point approximation \cite{Butler07,AmariWu}, and then Eq.~(\ref{eq:fin}) is obtained.
The goodness of the fit was evaluated estimating the normalised mean squared error (NMSE), $p-value < 0.05$.
It is then fitted the $q$ parameter to get the best fitting function $Q_{q}(r)$ in Eq.~(\ref{eq:fin}) for the experimental distribution. We used the matlab subroutine
GFIT2 to compute goodness of fit, for regression model, given matrix/vector of target and output values.
\subsection*{B. The Gaussian Distribution within the q-Information Geometry framework}
\setcounter{equation}{0}
\renewcommand{\theequation}{B-\arabic{equation}}

The q-Gaussian distribution is defined as:
\begin{equation}
p(x,\mu,\sigma)=\exp_{q} (-\frac{(x-\mu)^2}{2 \sigma^2} - \psi(\mu,\sigma) ),
\end{equation}
which can be rewritten as:
\begin{equation}
p(x,\mu,\sigma)=\exp_{q} (\frac{\mu}{\sigma^2} x - \frac{(x)^2}{2 \sigma^2} - \frac{(\mu)^2}{2 \sigma^2}- log_q (\sqrt{2 \pi}) ).
\end{equation}
where can identify the correlation coordinates as
\begin{equation}
\theta_1= \frac{\mu}{\sigma^2},
\end{equation}
\begin{equation}
\theta_2= \frac{-1}{ 2 \sigma^2},
\end{equation}
and the factor
\begin{equation}
\psi = \frac{(\mu)^2}{2 \sigma^2} + log_q (\sqrt{2 \pi} \sigma).
\end{equation}

\subsection*{C. Simulation}
\setcounter{figure}{0}
We applied our formalism to a network simulation model in which the number of neurons is a parameter
under control. In the following que consider a network of N = 200 neurons with the first
Ne = 80 of excitatory RS type, and the remaining Ni=120 of inhibitory FS
type \cite{Izhik03}. Each excitatory neuron is connected to M=80
random neurons, and each inhibitory neuron is connected to M=80 excitatory
neurons only. Synaptic connections among neurons have fixed conduction
delays of 10 ms. Notice that weak higher-order interactions of almost all orders are required for
realising a widespread activity distribution of a large population of neurons \cite{AmariWu}, this is
in agreement with the hypothesis of Amari a\cite{AmariWu} that the widespread probability distribution
in the thermodynamic limit is not realised even when pairwise interactions, or third-order interactions, exist
(see {Fig. ~\ref{infAp} }).

\newpage

\begin{figure}
\begin{center}
\includegraphics[width=6.in]{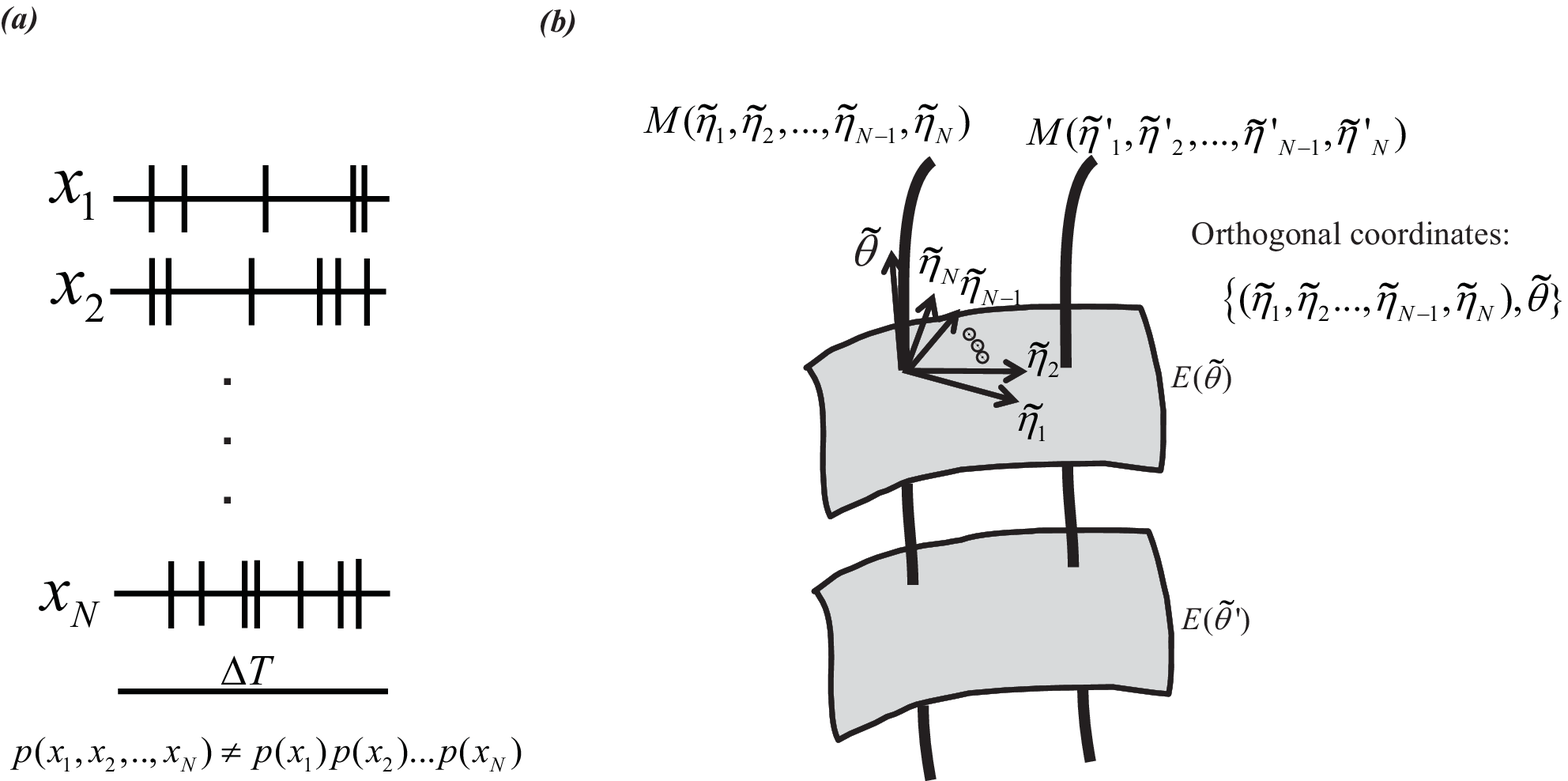}
\end{center}
\caption{\small{Schematic representation of spike correlations. (\emph{a}) Spikes being fired at a given time window of size $\Delta t$ considering a population of $N$ neurons.(\emph{b})$E(a_1)$, $E(a_2)$ and $E(a_3)$
are the family of distributions having the same correlation coordinates $a_1$, $a_2$, and $a_3$, respectively.
The family of all probability distributions belongs to the $q$-exponential family of distributions for any $q$, and thus we
can introduce the $q$-geometrical structure to any arbitrary family of probability distributions.}
\label{inf00}}
\end{figure}

\begin{figure}
\begin{center}
\includegraphics[width=6.in]{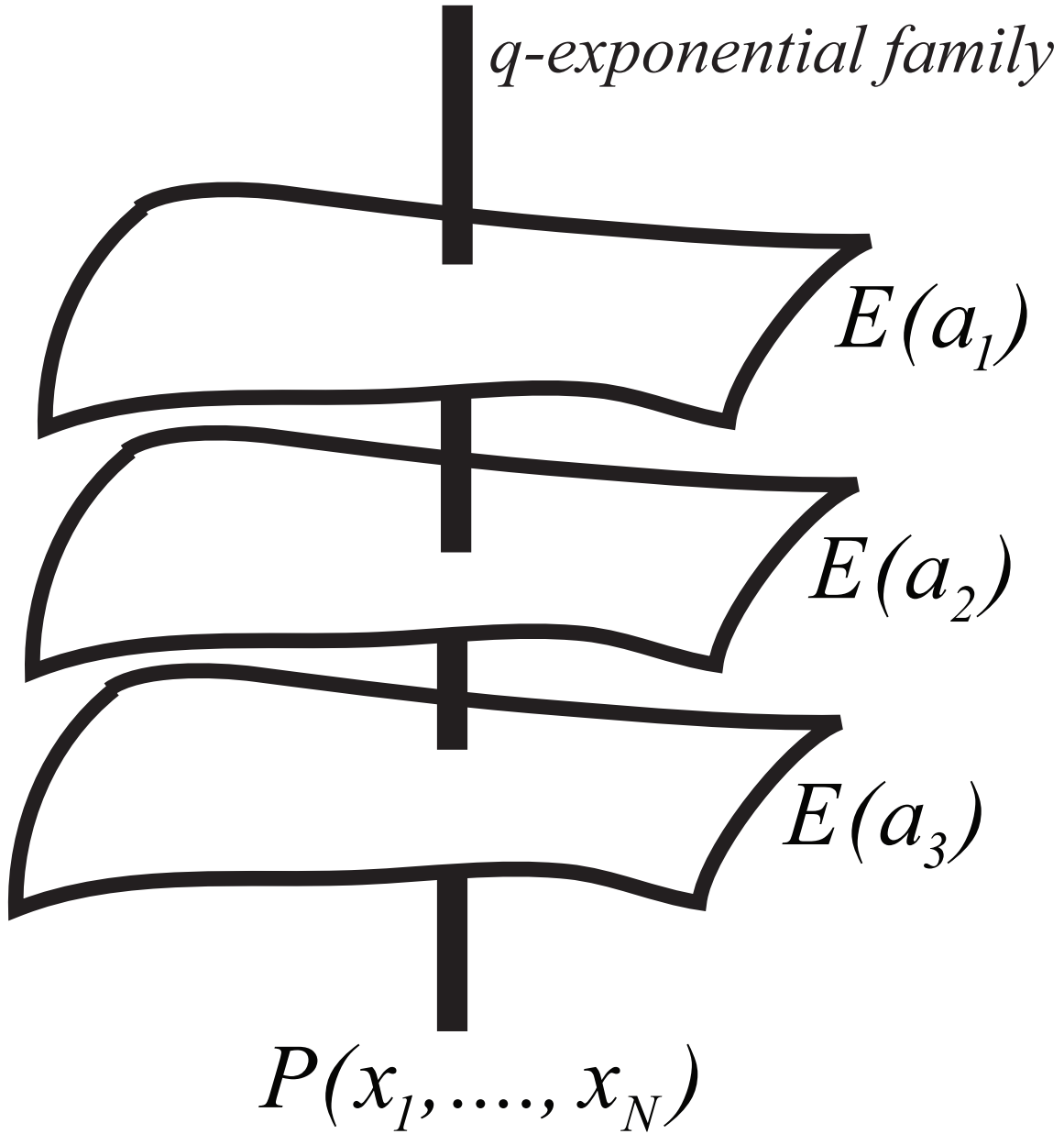}
\end{center}
\caption{\small{$E(a_1)$, $E(a_2)$ and $E(a_3)$
are the family of distributions having the same correlation coordinates $a_1$, $a_2$, and $a_3$, respectively.
The family of all probability distributions belongs to the $q$-exponential family of distributions for any $q$, and thus we
can introduce the $q$-geometrical structure to any arbitrary family of probability distributions.}
\label{inf00DD}}
\end{figure}

\begin{figure}
\begin{center}
\includegraphics[width=6.in]{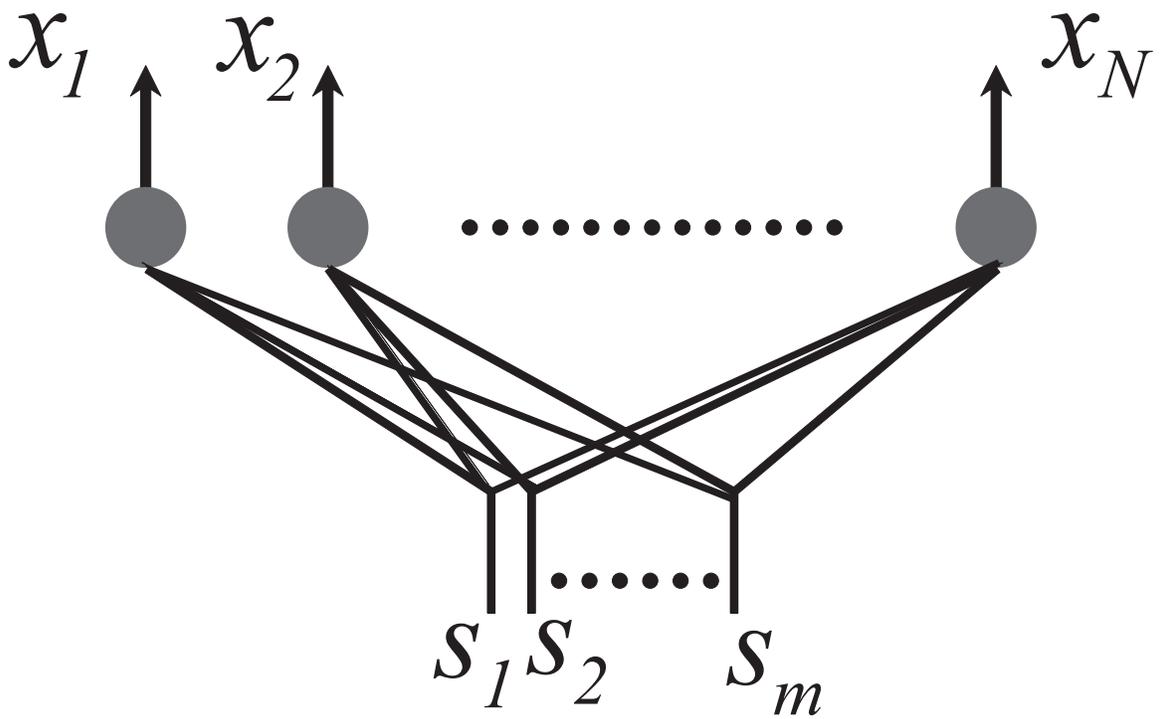}
\end{center}
\caption{\small{Schematic representation of a neuronal pool with $N$ neurons that receives $s_1,
s_2,...s_m$ common inputs.}
\label{inf0}}
\end{figure}

\begin{figure}
\begin{center}
\includegraphics[width=6.in]{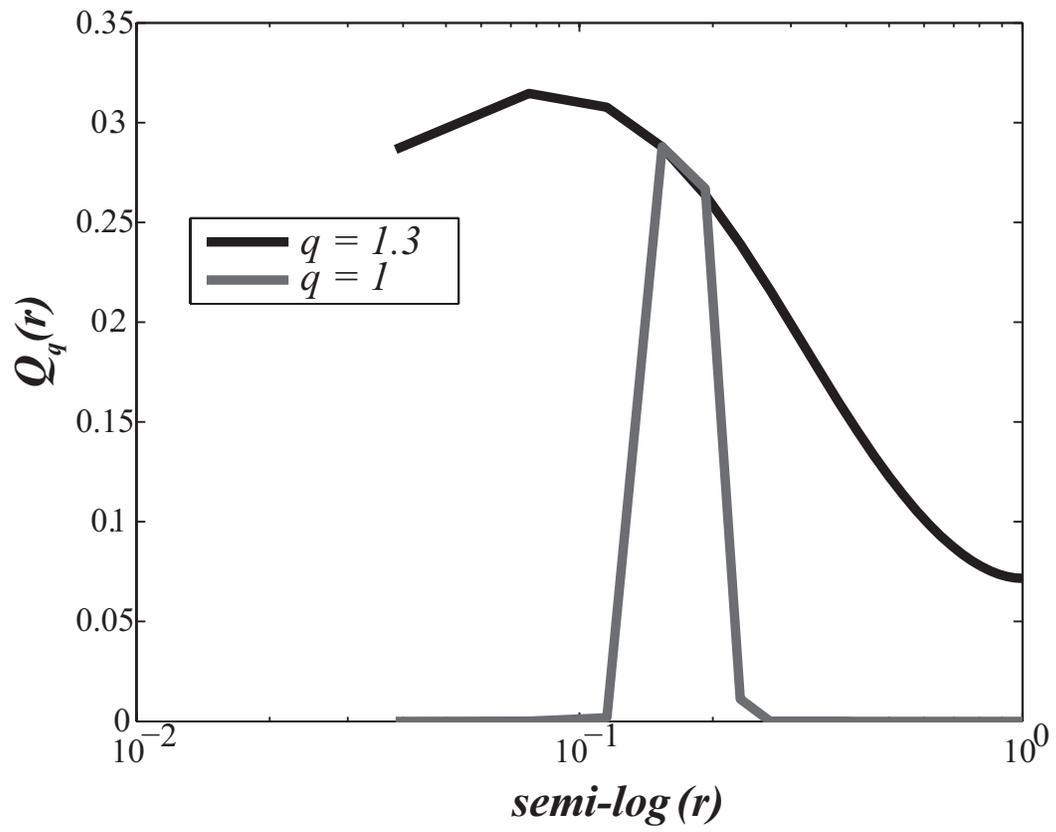}
\end{center}
\caption{\small{Distribution of firing $Q_{q}(r)$ computed for $q=1$ and $q=1.3$
(semi-log in the
X axis).}
\label{inf1a}}
\end{figure}

\begin{figure}
\begin{center}
\includegraphics[width=6.in]{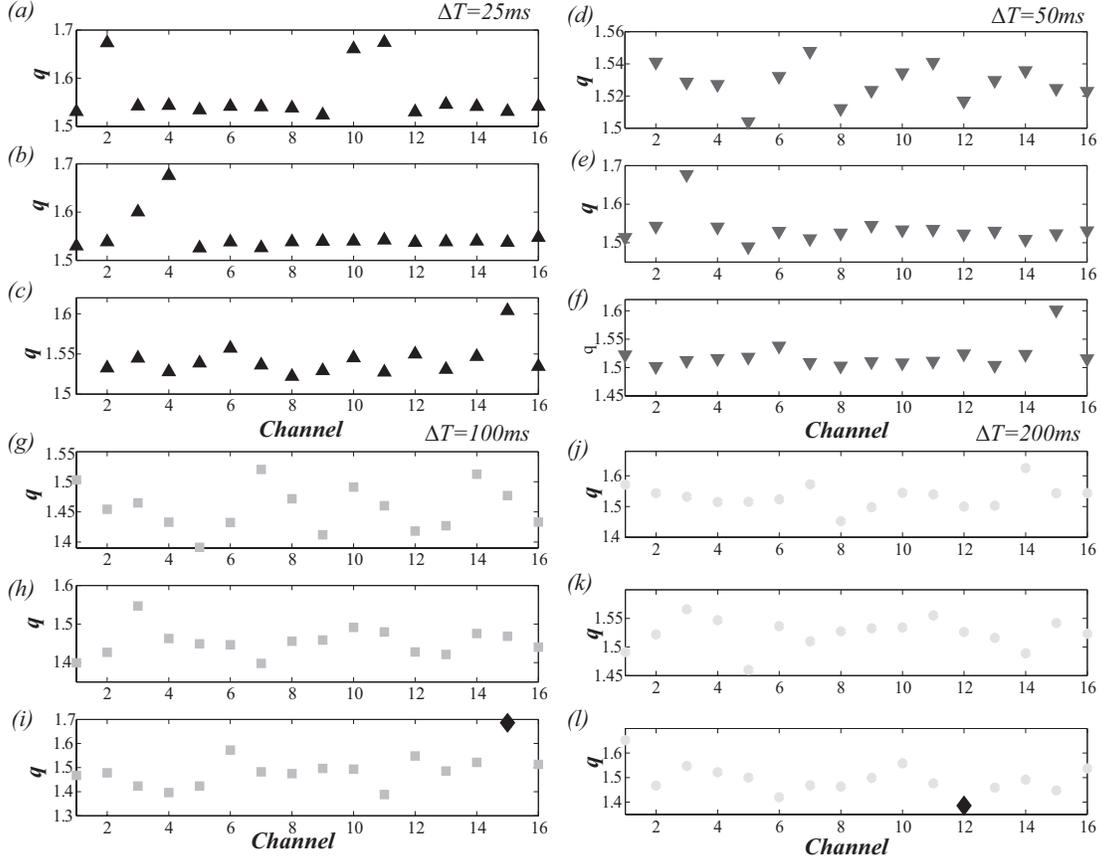}
\end{center}
\caption{\small{Parameter $q$ estimated through the fit of the firing distribution $Q_{q}(r)$ for the three different data sets.
 (\emph{a-c})  $\Delta T$=25 ms; (\emph{d-f}) $\Delta T$ = 50 ms;
 (\emph{g-i}) $\Delta T$ = 100 ms (maximum observed $q$: black diamond in panel \emph{i}, channel 15); and
 in (\emph{j-l}) the time windows is $\Delta T$ =200ms (minimum observed
$q$:  black diamond in panel \emph{l}, channel 12).
The goodness of the fit was evaluated by estimating the normalised mean squared error (NMSE), $p-value < 0.05$, see Appendix. }
\label{inf2}}
\end{figure}

\begin{figure}
\begin{center}
\includegraphics[width=6.in]{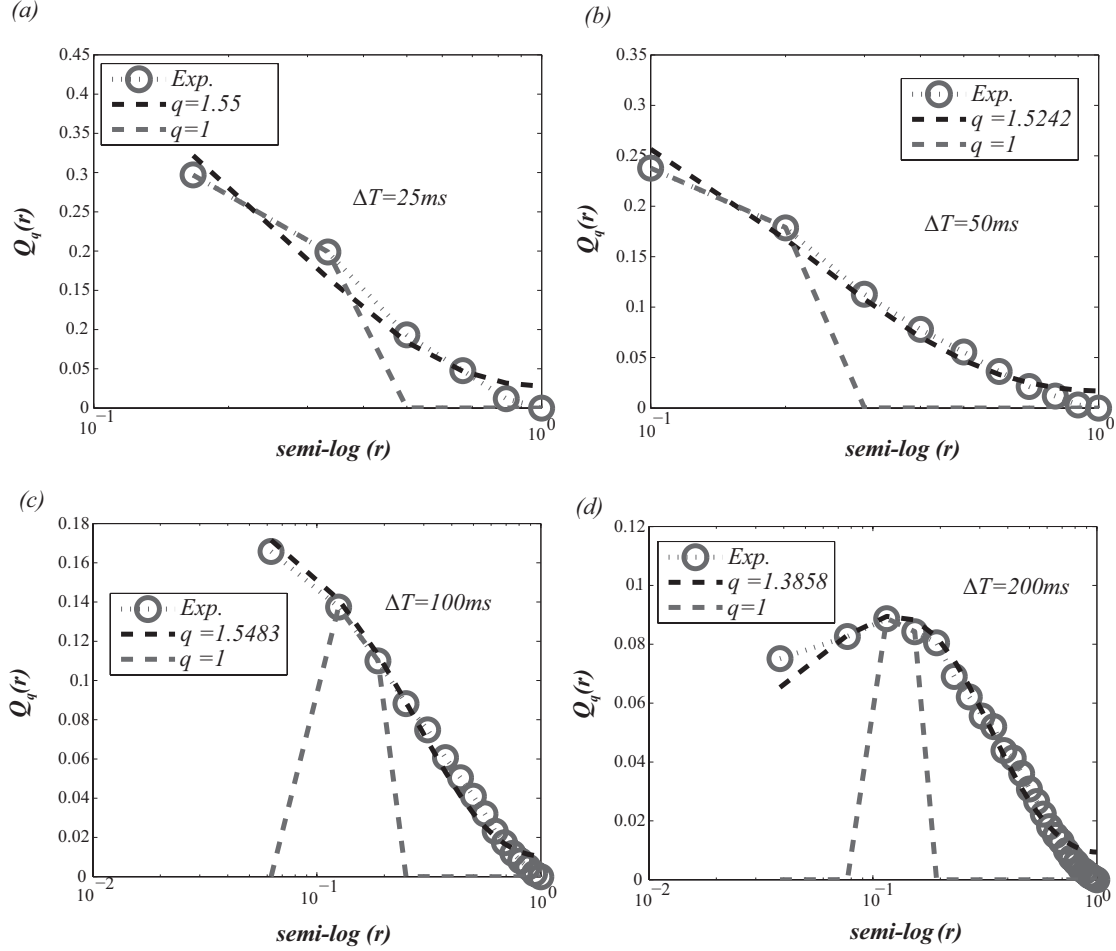}
\end{center}
\caption{\small{
Fit of $Q_{q}(r)$ as a function of the normalised firing rate $r$ (semi-log in the X axis). (\emph{a})  $\Delta T$=25 ms ($N_{max}=130$), $q=1.55$
and Fisher Info. equal to 0.1026 bits.
\emph{b}) $\Delta T$=50 ms ($N_{max}=160$), $q=1.5242$ and Fisher info. equal to 0.0489 bits.
(\emph{c}) $\Delta T$=100 ms ($N_{max}=200$),  $q=1.5483$ and Fisher info. equal to 0.0151 bits.
(\emph{d}) $\Delta T$=200 ms ( $N_{max}=240$),  $q=1.3858$ and Fisher info. equal to 0.0204 bits,
which corresponds to the minimum observed $q$ (Fig~\ref{inf2} panel \emph{l}: channel 12, diamond symbol).
The goodness of the fit was evaluated by estimating the normalised mean squared error (NMSE),  $p-value < 0.05$.
The theoretical approach  considering $q>1$ (black dashed line) shows a remarkably good fit,
in comparison to the case $q=1$ (grey dashed line), to the experimental curves (grey circles joined by dotted lines).
}
\label{inf3}}
\end{figure}

\begin{figure}
\begin{center}
\includegraphics[width=6.in]{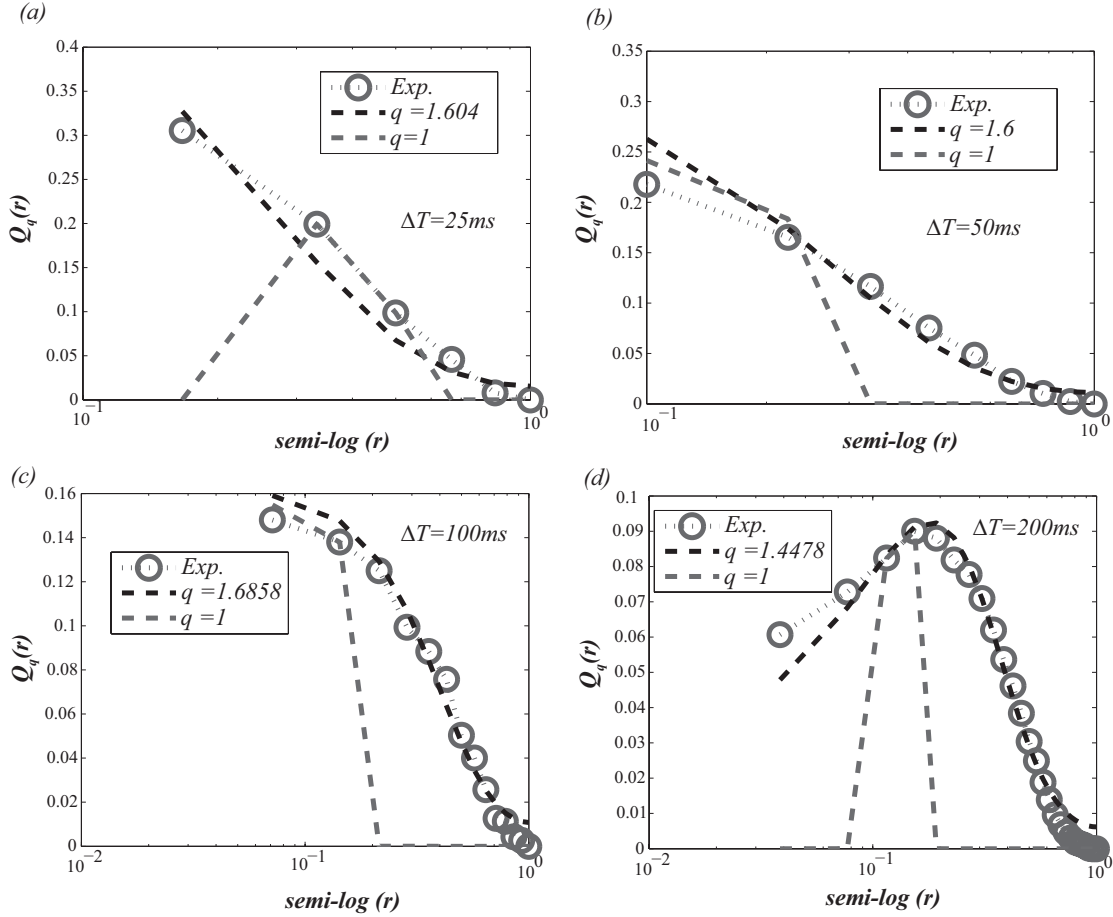}
\end{center}
\caption{\small{Fit of $Q_{q}(r)$ as a function of the normalised firing rate $r$ (semi-log in the X axis).
 (\emph{a})  $\Delta T$=25 ms ($N_{max}=130$), $q=1.604$ and Fisher info. equal to 0.0905 bits.
 (\emph{b}) $\Delta T$=50 ms ($N_{max}=140$),  $q=1.6$ and Fisher info. equal to 0.0395 bits.
 (\emph{c}) $\Delta T$=100 ms ($N_{max}=180$),  $q=1.6858$ and Fisher info. equal to 0.0142 bits,
 which corresponds to the maximum observed $q$ (Fig~\ref{inf2} panel \emph{i}: channel 15, diamond symbol).
 (\emph{d}) $\Delta T$=200 ms ($N_{max}=240$ ),  $q=1.4478$ and Fisher info. equal to 0.0150 bits. Same curve
labels as in Fig~\ref{inf3}.
}
\label{inf4}}
\end{figure}

\begin{figure}
\begin{center}
\includegraphics[width=6.in]{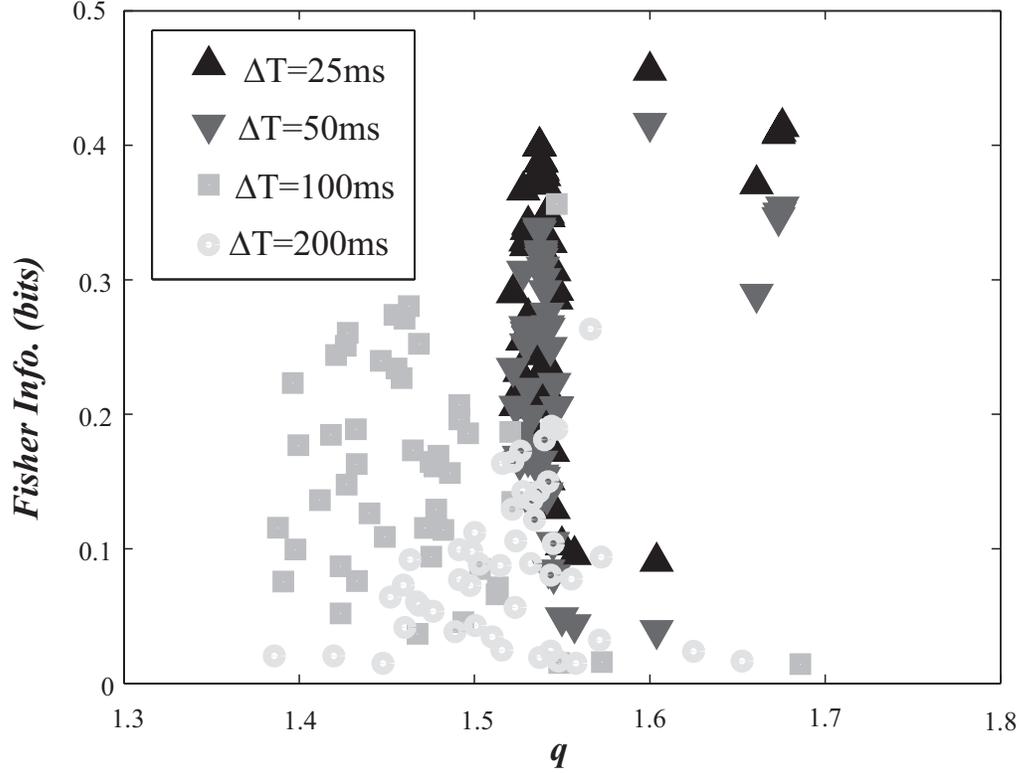}
\end{center}
\caption{\small{Fisher information versus $q$ (theoretical approach considering $q>1$),
taking four different time windows and considering all the 48 available channels.
Black up-triangles,  $\Delta T$=25 ms; dark down-triangles, $\Delta T$=50 ms;
dark grey squares, $\Delta T$=100 ms; light grey circles, $\Delta T$= 200 ms.
}
\label{inf5}}
\end{figure}

\begin{figure}
\begin{center}
\includegraphics[width=6.in]{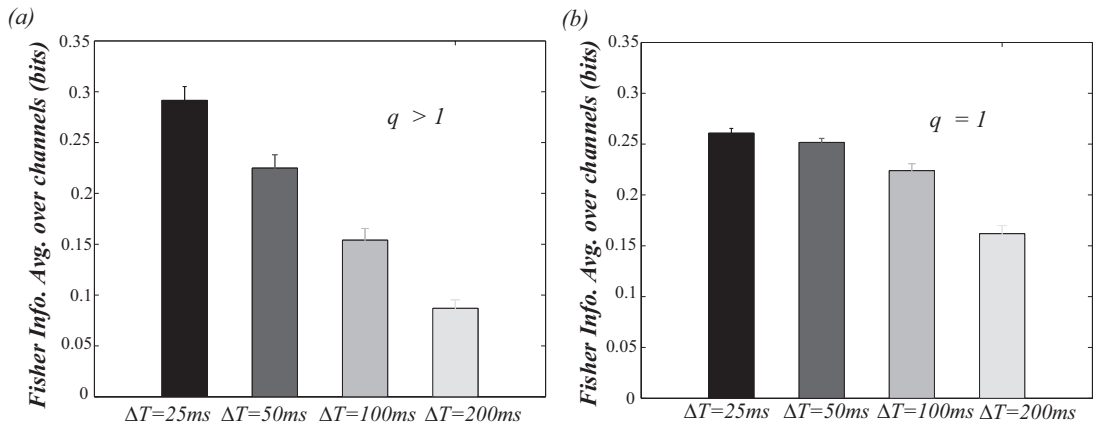}
\end{center}
\caption{\small{(\emph{a}) Fisher information averaged over the 48 different channels (theoretical approach considering $q>1$). Black bar,
$\Delta T$=25 ms; dark grey bar, $\Delta T$= 50 ms; grey bar, $\Delta T$=100ms; light grey bar, $\Delta T$=200ms.
(\emph{b}) Same as in B but considering the fit with $q$ constrained to be equal to 1.
}
\label{inf5a}}
\end{figure}

\begin{figure}
\begin{center}
\includegraphics[width=6.in]{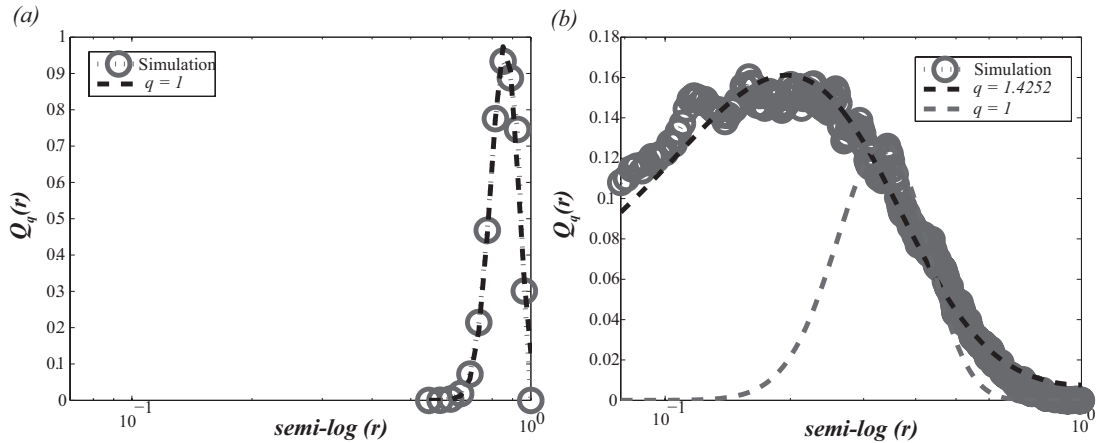}
\end{center}
\caption{\small{Distribution of firing $Q_{q}(r)$ versus the normalised firing rate $r$ (semi-log in the X axis), using 10 min of a simulated network
that consists of $N = 1000$ neurons with the first
$Ne = 650$ of excitatory RS type, and the remaining $Ni=350$ of inhibitory FS ($\Delta T$=25 ms).
(\emph{a}) Considering that each excitatory/inhibitory neuron is connected to $M=2$ or $3$ random neurons.
We find no difference in the probability distribution of firing when $M=2$ or $M=3$ is considered, $q =1$.
(\emph{b}) Considering that
the number of randomly interconnected neurons is $M=80$, $q=1.4252$.
 The goodness of the fit was evaluated by estimating the normalised mean squared error (NMSE),  $p-value < 0.05$.}
\label{inf00Da}}
\end{figure}

\begin{figure}
\begin{center}
\includegraphics[width=6.in]{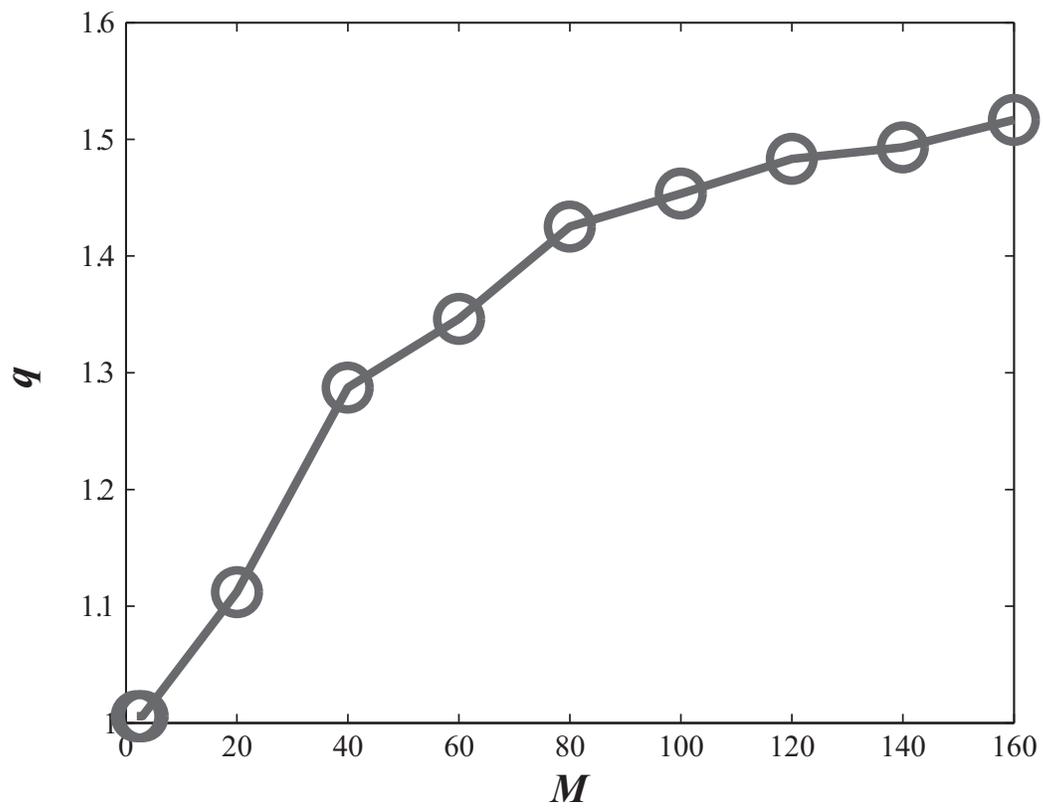}

\end{center}
\caption{\small{Parameter $q$ estimated through the fit of the firing distribution $Q_{q}(r)$,
versus the number of interconnected neurons $M$.
 The goodness of the fit was evaluated estimating the normalised mean squared error (NMSE),  $p-value < 0.05$.}
\label{inf00Db}}
\end{figure}

\setcounter{figure}{0}
\renewcommand{\thefigure}{C.1}
\begin{figure}
\begin{center}
\includegraphics[width=6.in]{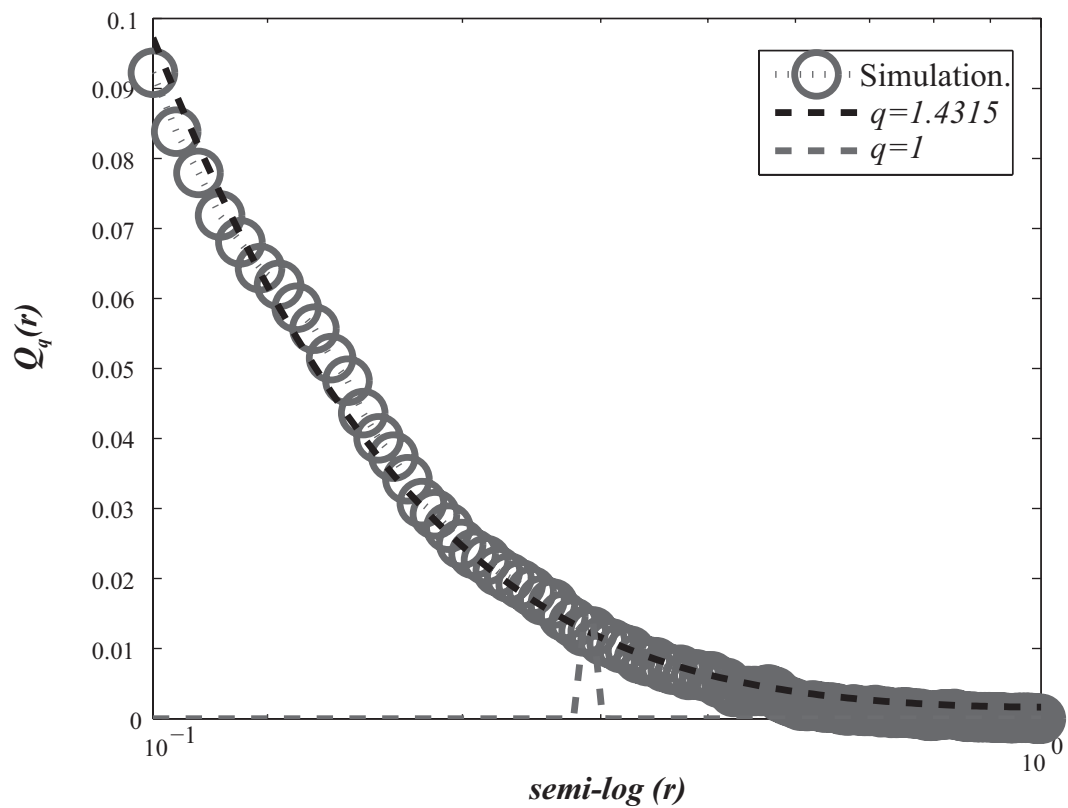}
\end{center}
\caption{\small{Distribution of firing $Q_{q}(r)$ using 10 min of simulated network
which consists of N = 200 neurons with the first
Ne = 80 of excitatory RS type, and the remaining Ni=120 of inhibitory FS ($\Delta T$=25 ms).
Considering that
the number of randomly interconnected neurons is M=80, $q=1.4315$.
 The goodness of the fit was evaluated estimating the normalised mean squared error (NMSE),  $p-value < 0.05$.}
\label{infAp}}
\end{figure}

\end{document}